\documentclass[a4paper,fleqn,usenatbib]{mnras}

\usepackage{mathptmx}

\usepackage[T1]{fontenc}
\usepackage{ae,aecompl}

\usepackage{graphicx}	
\usepackage{amsmath}	
\usepackage{amssymb}	
\usepackage{xspace, multirow,xcolor}
\usepackage{pbox}
\usepackage{lscape}
\usepackage{array}
\usepackage{float}
\usepackage{longtable}

\newcommand{\fig}[1]{Fig.~\ref{fig:#1}}

\newcommand{\tab}[1]{Table~\ref{tab:#1}}

\newcommand{\sect}[1]{Sect.~\ref{sec:#1}}
\newcommand{\app}[1]{Appendix~\ref{app:#1}}

\newcommand{\atlas}{ATLAS$^{3\mathrm{D}}$\xspace}
\newcommand{\rhoio}{\ensuremath{\rho_{10}}\xspace}
\newcommand{\mjam}{\ensuremath{M_\mathrm{JAM}}\xspace}

\title[LSB structures in MATLAS ETGs]{Census and classification of low-surface-brightness structures in nearby early-type galaxies from the MATLAS survey}

\author[M. B\'ilek et al.]{Michal B\'ilek$^{1}$\thanks{E-mail: bilek@astro.unistra.fr (MB)},
Pierre-Alain Duc$^{1}$,
Jean-Charles Cuillandre$^{2}$,
Stephen Gwyn$^{3}$,\newauthor
Michele Cappellari$^{4}$,
David V. Bekaert$^{5}$,
Paolo Bonfini$^{6}$,
Theodoros Bitsakis$^{7}$,\newauthor
Sanjaya Paudel$^{8}$,
Davor Krajnovi\'c$^{9}$,
Patrick R. Durrell$^{10}$,
Francine Marleau$^{11}$
\\
$^{1}$ Universit\'e de Strasbourg, CNRS, Observatoire astronomique de Strasbourg (ObAS), UMR 7550, 67000 Strasbourg, France\\
$^{2}$ IRFU, CEA, Universit\'e Paris-Saclay, Universit\'e Paris Diderot, AIM, Sorbonne Paris Cit\'e, CEA, CNRS, Observatoire de Paris, \\ PSL Research University, F-91191 Gif-sur-Yvette Cedex, France\\
$^{3}$ NRC Herzberg Astronomy and Astrophysics, 5071 West Saanich Road, Victoria, BC, V9E 2E7, Canada\\
$^{4}$ Astrophysics, University of Oxford, Denys Wilkinson Building, Keble Road, Oxford, OX1 3RH, England\\ 
$^{5}$ MCG Department, Woods Hole Oceanographic Institution, Woods Hole, MA, USA\\
$^{6}$ University of Crete -- Department of Physics, Voutes University Campus, GR-71003 Heraklion, Greece\\
$^{7}$ Institute of Astrophysics, FORTH, GR-71110 Heraklion, Greece\\
$^{8}$ Department of Astronomy and Center for Galaxy Evolution Research, Yonsei University, Seoul 03722, Korea\\
$^{9}$ Leibniz-Institut f\"ur Astrophysik Potsdam (AIP), An der Sternwarte 16, D-14482 Potsdam, Germany\\
$^{10}$ Department of Physics \& Astronomy, Youngstown State University, Youngstown, OH 44555 USA\\
$^{11}$ Institute of Astro and Particle Physics, Technikerstrasse 25/8, University of Innsbruck, A-6020 Innsbruck, Austria\\
}

\date{Accepted XXX. Received YYY; in original form ZZZ}

\pubyear{2015}

\begin{document}
\label{firstpage}
\pagerange{\pageref{firstpage}--\pageref{lastpage}}
\maketitle


\begin{abstract}
 The morphology of galaxies gives essential constraints on the models of galaxy evolution. The morphology of the features in the low-surface-brightness regions of galaxies has not { been} fully explored yet because of observational difficulties. Here we present the results of our visual inspections of very deep images of a  large volume-limited sample of 177 nearby massive early-type galaxies (ETGs) from the MATLAS survey. The images reach a surface-brightness limit of $28.5-29$\,mag\,arcsec$^{-2}$ in the $g'$ band. Using a dedicated navigation tool and questionnaire, we looked for structures at the outskirts of the galaxies such as tidal shells, streams, tails,  disturbed outer isophotes or { peripheral  star-forming disks}, and simultaneously noted the presence of contaminating sources, such as Galactic cirrus. We also inspected internal sub-structures such as bars and dust lanes. We discuss the reliability of this visual classification investigating the variety of answers made by the participants.  We present the incidence of these structures and the trends of the incidence with the mass of the host galaxy and the density of its environment. We find an incidence of shells, stream and tails of approximately 15\%, about the same for each category. For galaxies with masses over $10^{11}$\,M$_\odot$, the incidence of shells and streams { increases about 1.7 times}.  We also note a strong unexpected anticorrelation of the incidence of Galactic cirrus with the environment density of the target galaxy. Correlations with other properties of the galaxies, and comparisons to model predictions,  will be presented in future papers. 
 
%
%
\end{abstract}

\begin{keywords}
galaxies: elliptical and lenticular, cD -- galaxies: interactions -- galaxies: peculiar -- galaxies: haloes -- galaxies:photometry -- galaxies: structure
\end{keywords}



\section{Introduction}
{Deep galaxy imaging of nearby extended systems is a dynamically growing part of present-day observational astronomy. With current large-field-of-view cameras and dedicated observing techniques, we can get close to a surface-brightness limit of 30\,mag\,arcsec$^{-2}$ in a few hours of observations or even less. Deep imaging is being used for multiple applications. 



{For instance, it can be employed to detect tidal features, the remnants of galaxy interactions. They can be either disturbances made by tidal forces exerted by one or more close-by companions in an  on-going interaction, or be made of  material that originally belonged to another  galaxy. Tidal features are useful in many regards; for instance their incidence and morphology can then be compared to simulations to test galaxy formation theories   \citep[e.g.,][]{bullock05,johnston08,cooper10,hendel15,pop18,mancillas19}, tidal features can help us clarify how unusual galaxies were formed (e.g., \citealp{sanders96,draper12,koshy17,oh19,muller19c,ebrova20}) and they moreover are  useful probes of   gravitational fields \citep[e.g.,][]{ebrova12,sanderson13,bil13,bil15,bil15b,sanderson17,thomas17,thomas18,malhan19}, among others. }

Deep imaging also discloses the extended stellar halos \citep[e.g.,][]{trujillo13,trujillo16,merritt16,mihos17,iodice17b,rich19,iodice20} made of accreted material, i.e. dissolved tidal features, or that formed during the dissipational-collapse stage of galaxy assembly. In specific cases, faint extended star-forming disks may be revealed, even around passive early type galaxies \citep{duc15}. 

Detecting low-surface brightness galaxies on deep images became popular recently \citep{javanmardi16,greco18,mihos18,habas19,muller19a}. The nature and formation of these galaxies is an interesting question by itself \citep[e.g.,][]{dabring13,amorisco16,dicintio17,chan18,carleton19,liao19}; the dwarf galaxies also serve as test-beds for cosmological models \citep[e.g.,][]{kroupa05,muller16,libeskind16,banik18,pawlowski19,javanmardi20}. { Even though the first ultra-diffuse galaxies were discovered a long time ago \citep{sandage84,impey88,dalcanton97}, the new observing techniques helped to raise a strong interest of scientists in this type of galaxies \citep[e.g.,][]{vandokkum15,koda15,mihos15,roman17a,roman17b,papastergis17,vandokkum18a,vandokkum18b,bennet18,gonzalez18,alabi18,torrealba19,janssens19}.}

Finally, a fraction of the deep images capture galactic cirri, the diffuse part of the interstellar medium of our own galaxy. Although more prominent towards lower Galactic latitude, they are present in the whole sky and hinder the detection of low surface brightness objects behind them.  One may take advantage of them  studying  the processes in the ISM at spatial scales inaccessible before \citep{miville16,roman19}.

Deep images raise challenges at the observational and data-reduction stages, in particular related { to} the instrumental large-scale sky background  variations, ghost halos and other PSF effects. These effects produce artifacts that can either be confused with the real astronomical objects, or make the detection of the objects of interest uncertain. The pitfalls of deep imaging are illustrated by the recent controversy about the tidal streams of the Splinter Galaxy, NGC\,5907. Several amateur observers equipped by small telescopes have observed the streams warping the galaxy multiple times -- the refereed example is \citet{martinezdelgado08} -- while the professional astronomers with bigger telescopes  were able to detect only a part of this structure \citep{laine16,vandokkum19,muller19b,alabi20}. The case remains unresolved.

Once the images are reduced, we have to identify the structures of interest. Let us present the methods on  the example of tidal features,  the structures that have probably been investigated most in the past. The vast majority of existing works rely, at least partly, on the visual inspection of images whose brightness scale has been adjusted suitably \citep[e.g.,][]{bridge10, kaviraj10, nair10,sheen12, atkinson13}. Others employ  more elaborate methods that may include removing point-like sources and smoothing  the image by a Gaussian or a median kernel filter \citep[e.g.,][]{tal09, miskolczi11, adams12,  morales18, hood18}, unsharp masking technique \citep[see the pioneering work by][]{MC83} and  subtraction of galaxy models \citep{mcintosh08} that  enhance the visibility of faint streams and shells. For detecting tidal features in large galaxy samples, automated methods have to be employed.  \citet{kadofong18} preselected galaxies for visual inspection using an algorithm based on detecting high spatial frequencies in the image. Another method relies on the tidal parameter, a function of the average ratio of the original image and a smooth model of the galaxy \citep{vandokkum05,adams12}. \citet{pawlik16} developed a method based on the asymmetry of isophotes. The  coefficients such as concentration, asymmetry, Gini or $M_{20}$  \citep{abraham96, conselice03, lotz04} work well for detecting perturbed galaxies at high redshift, nevertheless they are not suitable for detecting tidal features  whose luminosity comprises just a small fraction of the total luminosity of the galaxy. We do not { just wish  to  know} if tidal features are present --  their morphological type and number provide further precious { constraints}. The automatic algorithms listed so far cannot do that in contrast with the visual inspection. Promising methods for both detecting and classifying galaxy substructures are convolutional neural networks \citep{walmsley18, pearson19}, that reach an $\sim80\%$ completeness with a $\sim20\%$  contamination or the algorithm called ``Subspace-Constrained Unsupervised Detection of Structure'' \citep{hendel19}. The importance of the automated methods will grow with the advent of the unprecedented number of new images produced by the large future surveys, see \sect{future}.

The price to pay for  deep imaging is the requirement of  relatively long observing time coupled with the need to observe large number of systems. A number of observing efforts have been  performed to achieve this.  We compiled the characteristics of some of the ongoing, recent or notable surveys looking for tidal features or exploring low-surface brightness structures in Table \ref{tab:surveys}. 
Note that we have not included in this list the limiting surface brightness. Indeed, different authors use different methods to estimate it making comparisons difficult. Differences may be as large as 2--3 mag from one method to the other, depending on whether integrated profiles or local measurements are made. {Instead, we state whether the survey was optimized for detecting extended low-surface-brightness structures, i.e. that the large-dithering strategy was employed (see, e.g., \citealp{duc15}).}

In our work we exploit deep optical images of a complete volume-limited sample of 177 nearby ETGs taken in the MATLAS survey \citep{duc15,habas19}. All galaxies studied here are a part of the \atlas reference sample \citep{cappellari11a} meaning that a lot of additional information is available about them and hence it will be possible to study the relations of the faint structures with other properties of the galaxies. In addition, the originality of the MATLAS survey, compared to similar studies,  lies in the \textit{combination} of depth, LSB optimization, the number of target galaxies, detailed classification methods and image quality. The excellent seeing at CFHT and the large diameter of the telescope provide images having a much better angular resolution than $<1\,m$ telescopes.

In this paper we present a catalogue of the following types of structures/features/objects in or around the MATLAS ETGs: 1) tidal features (streams, shells, tails), 2) disturbed outer isophotes, 3) stellar bars and other features formed by secular evolution of rotating galaxies, 4) dust lanes, 5) { peripheral star-forming disks}, and finally 6) galactic cirri.  We investigate their incidence, and for the tidal features also their number in the galaxy. Finally, we explore how the incidence correlates with the mass and environmental density of the galaxies,  which are important properties influencing the evolution of galaxies. Since our work will be a basis for subsequent works within the \atlas project, we provide a detailed discussion of the possible biases of the method we used for our structure detection and classification. }

The paper is organized as follows. In \sect{matlas} we give details about our sample and the MATLAS survey. Our methods are explained in \sect{meth} and the results are presented in \sect{res}. Section~\ref{sec:disc} is the discussion where we consider the advantages and shortcomings of the method (\sect{discmeth}), the comparison of our results with literature (\sect{lit}) and provide a qualitative interpretation of the results (\sect{interp}). Section~\ref{sec:future} deals with our future plans. We summarize in \sect{sum}. { Supplementary information is given in the appendices. Our methods to classify the galaxies are described more in detail in \app{meth}. In \app{experience}, we explored how  experience of the participants affects their classifications of the galaxies. Finally, in \app{img} we provide example images of the various types of features of interest for this paper.}

\newcolumntype{P}[1]{>{\raggedright\arraybackslash}p{#1}}
{
\renewcommand{\arraystretch}{1.3}
\begin{table*}
	\caption{Comparison of MATLAS to other surveys targeting tidal features or low-surface-brightness objects.}
	\label{tab:surveys}
	\centering
	\begin{minipage}{\textwidth}
	\begin{tabular}{P{28mm}P{10mm}lP{15mm}P{18mm}P{6mm}P{21mm}P{10mm}P{15mm}}
		\hline\hline
Paper (survey name) & \pbox{13mm}{LSB optimized?} & \pbox{9mm}{ Coverage \newline [deg$^2$]} & Bands  & Telescope  & \pbox{15mm}{FOV$^\mathrm{g}$ \hspace{5mm}[deg$^2$]} & \pbox{20mm}{Targets} & \pbox{13mm}{Number \hspace{1mm} of objects} & \pbox{15mm}{Distance [Mpc]}\\\hline
This work (MATLAS) & Y & 144  & $(u^*) g^\prime r^\prime (i^\prime)^\mathrm{f}$ & 3.6m CFHT & 1 & ETGs & 177 & <40\\
\citet{annibali20} (SSH) & Y & 6.6 & $gr$ & 11.9m LBT & 0.15 & LTG dwarfs & 45 &  $\lesssim 10$\\
\citet{danieli19} (DWFS$^\mathrm{a}$) & Y & 330$^\mathrm{d}$ &  $gr$ & 1m Dragonfly & blind & -- & -- & -- \\
\citet{gilhuly19} (DEGS$^\mathrm{a}$) & Y & -- &  --& 1m Dragonfly & 4.9  & Edge-on LTGs & -- & -- \\
\citet{rich19} (HERON$^\mathrm{a}$) & Y & 68 &  $L$ & 0.7m Jeanne Rich  & $0.57$ & Nearby gxs. & 119 & <50 \\
\citet{hood18} (RESOLVE) & N+Y & 710 & $r$ & 4m CTIO + 2.5m Sloan & blind & $M_{*+\mathrm{HI}}$ $\gtrsim 10^{9.2}$\,M$_\odot$  & 1048 & $\sim60-100^\mathrm{b}$\\
\citet{byun18} (KMNet$^\mathrm{a}$) & Y  & -- & -- & 1.6 KMNet & 4 & -- & -- & -- \\
\citet{kadofong18} (HSC-SSP$^\mathrm{a}$) &  N & $200^\mathrm{c}$/$1400^\mathrm{d}$ & $i^\mathrm{c}$ & 8.2m Subaru & blind & all & 21208 & 200-2500$^\mathrm{b}$ \\
\citet{morales18} (SDSS) & N & 74.3 & $g+r+i$  & 2.5m Sloan & blind  & $M_*=10^{10\mathrm{-}11}$\,M$_\odot$ & 297 & $\lesssim 30$\\
\citet{stss} (STSS$^\mathrm{a,e}$) & -- & -- & $L$ & 0.1-0.5m amateurs &  >0.25 & $M_K<-19.6$ & -- & <40\\
\citet{peters17} (SDSS Stripe 82$^\mathrm{i}$) & Y & 275  & $ugriz$ & 2.5m Sloan & blind & Face-on LTGs & 22 & mostly<100 \\
\citet{mihos17} (BSDVS) & Y & 16 & $BV_M$ & 0.61m Burrell Schmidt & blind & Virgo Cluster & --  & -- \\
\citet{iodice16,iodice17} (FDS) & Y & 26 & $ugri$ & 2.6m VST & blind & Fornax Cluster & -- & 20 \\
\citet{munoz15} (NGFS$^\mathrm{h}$) & Y & 30 & $u^\prime g^\prime r^\prime i^\prime JK_S $ & 4m CTIO & blind & Fornax Cluster & -- & -- \\
\citet{capaccioli15} (VEGAS$^\mathrm{a}$) & Part & $\sim 100$ & $gri$ & 2.6m VST & 0.9 & ETGs $B<-19.2$  & $\sim 100$ & $<60^\mathrm{b}$ \\
\citet{atkinson13} 	(CFHTLS) & N & 170 & $g^\prime+r^\prime+i^\prime$  & 3.6m CFHT &  1 & $r^\prime<15.5$ ($M_{r^{\prime}}<19.3$) & 1781 & $180-690^\mathrm{b}$\\
\citet{Ferrarese12} (NGVS) & Y & 104 & $u^*g(r)iz^\mathrm{f}$ & 3.6m CFHT & 1 & Virgo Cluster & -- & -- \\
\citet{adams12} (MENeaCS) & N & 54 & $r$ & 3.6m CFHT & 1 & ETGs $M_r<-20$ & 3551 & 180-720$^\mathrm{b}$ \\
\citet{sheen12} & N & $\sim1.5$ & $ugr$ & 4m CTIO & blind & red cluster-gxs. $M_r<-20$  & 273 & 200-520$^\mathrm{b}$ \\
\citet{miskolczi11}	(SDSS DR7) & N & 8423 & $g+r+i$ & 2.5m Sloan & blind & edge-on LTGs & 474 & mostly < 100\\
\citet{kaviraj10}(SDSS Stripe 82) & N & 270 & $r$ & 2.5m Sloan & blind & ETGs $M_r<-20.5$ & $\sim 300$ & $\lesssim 220^\mathrm{b}$\\
\citet{nair10}  (SDSS DR4) & N & 6670 & $g^\prime r^\prime i^\prime$  & 2.5m Sloan & blind & $g$<16 & 14034 & $40-460^\mathrm{b}$\\
\citet{bridge10} (CFHTLS-Deep) & N & 2 & $u^* g^\prime r^\prime i^\prime z^\prime$ & 3.6m CFHT  & blind & $i_{Vega}\leq22.2$ & 27000 & $990-84000^\mathrm{b}$\\
\citet{tal09} (OBEY) & Y & $\sim 6$ & $V$ & 1m CTIO & 0.11 & Es, $M_B<-20.15$ & 55 & 15-50 \\
\citet{vandokkum05} (MUSYC, NDWF) & N & 10.5 & $UBVRIz$, $B_wRI$ & 4m Mayall, 4m CTIO &  blind & ETGs & 126 & mean $7500^\mathrm{b}$\\
\citet{MC83} & N & -- & IIIa-J emulsion & 1.24m UK Schmidt & 44 & ETGs & 137 & majority < 300 \\

		\hline
	\end{tabular}
\end{minipage}
\begin{minipage}{\textwidth}
\footnotesize{\textbf{Notes}: $^\mathrm{a}$ Ongoing surveys. $^\mathrm{b}$ Estimated from redshift assuming $H_0 = 69.6$\,km\,s$^{-1}$, $\Omega_\mathrm{m} = 0.286$ and flat cosmology as the luminosity distance using the calculator of \citet{wright06}. $^\mathrm{c}$ Used in the paper. $^\mathrm{d}$ Intended final coverage. $^\mathrm{e}$ See also \citet{martinezdelgado09}. $^\mathrm{f}$ Images in the bands listed in parenthesis were available only for a part of the investigated galaxies. $^\mathrm{g}$ The ``blind'' fields mean that the target is an area of the sky, not a particular object.  $^\mathrm{h}$ See also \citet{eigenthaler18}. $^\mathrm{i}$ See also \citet{fliri16}.}
\end{minipage}
\end{table*}
}

\begin{table}
\caption{Survey characteristics {of MATLAS}.}
\label{tab:survey}
\begin{tabular}{ccccc}
\hline\hline
& $u^*$ & $g^\prime$ & $r^\prime$ & $i^\prime$ \\ \hline
Number of observed galaxies & 12 & 178 & 179 & 104 \\
Average seeing $\left[ \rm{arcsec} \right]$ & 1.03 & 0.96 & 0.84 & 0.65 \\ \hline
\end{tabular}
\end{table}

\begin{figure*}
	\centering
	\includegraphics[width=17cm]{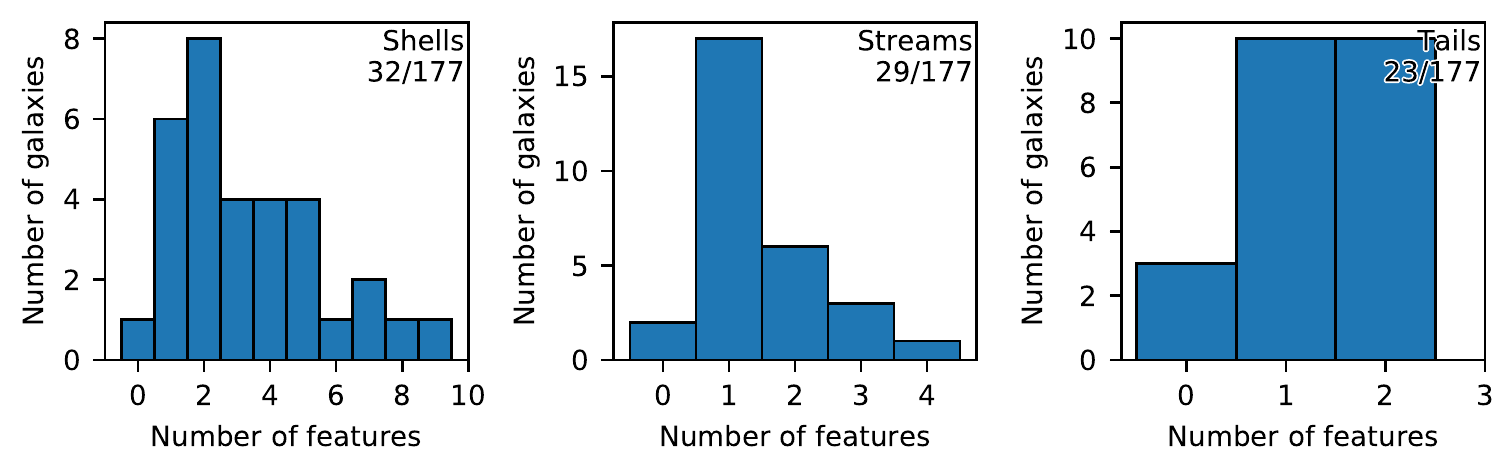}
	\caption{Histograms of the number of shells (left), streams (middle) and tails (right) per galaxy in our sample. Here are included only the galaxies where the number of detected features was greater than zero (the galaxies in the zero bin have, e.g., 0.5 streams, i.e. one likely stream, see \app{summarizing}). The number of galaxies that were counted in in these histograms and the total number of  galaxies in the sample are noted in the corners.}
	\label{fig:freqhist}
\end{figure*}

\section{The MATLAS survey}
\label{sec:matlas}

The images used in this paper were obtained as part of the MATLAS deep imaging project carried out at the 3.6\,m Canada-France-Hawaii Telescope (CFHT) using the MegaCam imager. The instrument combines the light gathering power of a relatively big telescope with a wide field-of-view. The images reach local surface brightness limit of 28.5-29\,mag\,arcsec$^{-2}$ in the $g^\prime$ band \citep{duc15}. The large field-of-view of $1\times1^\circ$ allows detecting extended tidal structures around nearby galaxies as well as inspecting their environment, including the presence of contaminating cirrus emission at large scales. The latter usually appears as parallel stripes, which might be confused with tidal features if the field-of-view is not large enough.

The MATLAS project  developed as a CFHT Large Program  made in the framework of the \atlas collaboration \citep{cappellari11a}. As such,  it  benefited from the availability of the  multi-wavelength spectroscopic and imaging observations with various instruments and telescopes (Sauron/WHT, Westerbork radio telescope, IRAM, etc.). The acquisition of the deep  images  was completed by short-exposure observations of the galaxies whose inner regions were saturated. The details are described in \citet{duc15}, together with preliminary results based on the observations of a fraction of the sample presented here. An exhaustive description of the full  MATLAS survey, including the field positions and observing conditions, is presented in \cite{habas19}. Table~\ref{tab:survey} summarizes the main characteristics of the survey. 

 The reference sample for this study, the \atlas sample with 260 members,  is characterized in \citet{cappellari11a}. Briefly, it is a complete volume-limited sample of  nearby galaxies (distance below 42 Mpc), that are massive (absolute $K$-band magnitude below -21.5), easily observable from the northern hemisphere ($|\delta-29^\circ|<35^\circ$), that avoid the obscuration by the Milky Way ($|b|>15^\circ$) and were visually classified as early-type, based on shallow optical images from the Sloan Digital Sky Survey \citep{sdss7} or the Isaac Newton Telescope.  The MATLAS sub-sample was obtained from the \atlas sample by excluding the 58 galaxies belonging to the Virgo cluster, since they have already had  deep images from the ``Next Generation Virgo Cluster Survey'' \citep{Ferrarese12} when the project was conceived. Therefore  MATLAS    does not probe the densest environments since all \atlas galaxies in clusters reside in Virgo.  An analysis similar as the one presented here for the \atlas galaxies in Virgo is planned.  Furthermore, the MATLAS sample does not include   galaxies that are located close to bright stars as they are unsuitable for deep imaging (about 20 of them). 
 The final MATLAS sample analyzed in the present paper includes  177  ETGs for which at least two bands ($g'$ and $r'$) are available. Additional $i'$ band and $u^*$ band images were obtained for resp. 60\% and   7\% of this sample.

The primary objective of MATLAS is the systematic census of the relics of past collisions, i.e.  tidal features  and extended stellar halos. While the first type of structures is discussed in the current paper, exploiting the images of stellar halos requires corrections for the light scattered by the optical elements of the camera, and  the wide wings of the PSF.  A deconvolution technique is used for this \citep{Karabal17}. Results will presented in  Y{\i}ld{\i}z et al. \textit{in prep.}. The  study of diffuse extended star-forming disks traced by blue and UV emission or dust lanes has been detailed in \citet{Yildiz17} and \citet{Yildiz20}. The census of such relatively rare features around ETG is also presented in the present paper. 

Because of the large field-of-view of the MegaCam camera,  the MATLAS survey allows studying the large-scale environment of the target ETGs: massive spiral companions (about 100 of them), but also dwarf galaxy satellites, including ultra-diffuse galaxies, { and associated globular clusters (GCs), which are also good tracers of the past assembly of galaxies. \citet{habas19}  described a sample of about 2200 dwarf candidates from the MATLAS images.  A catalog of GC candidates in MATLAS images was compiled by Lim et al. \textit{in prep}. The association between GCs and some collisional features for one system is presented in \citet{Lim17} and Fensch et al. 2020 (\textit{submitted}).  In principle, it will be possible to estimate the dynamical mass  of the  galaxies in our sample from the number of their GCs \citep[e.g.,][]{harris15,forbes18} or kinematics \citep[e.g.,][]{samur14,bil19} after measuring the radial velocities of the GCs. In this paper  however, our mass dependence analysis makes use of a proxy of the stellar mass.  }

A fraction of the MATLAS images capture Galactic cirri, dust clouds in our own Milky Way, some located close to the Sun. \citet{miville16}  illustrated the great scientific potential of such images for studying the turbulence cascade in the interstellar medium at high spatial resolution. Fields affected by cirrus are systematically  annotated in this paper.

\section{Methods}
\label{sec:meth}
{We give here a brief description of our fine structure identification and classification methods; they are detailed in \app{meth}.  The results of this paper are based on visual inspection of the MATLAS images. The images were { inspected} by six scientists, all specialists in observing or simulating nearby galaxies. Each of them classified at least two-thirds of the galaxies, and the majority all of them.
Clearly, the number is low compared to  citizen-science projects such as Galaxy Zoo \citep{galaxyzoo}, but this is compensated by a much stronger expertise of our classifiers.   
In fact, the initial group of participants  included more people (15). An analysis of the votes had shown a decrease of the degree of consensus when the votes of the less experienced classifiers were included.

The participants were provided an online navigation tool for displaying the images  that allowed them to navigate through the images, zoom in and out, change the luminosity scaling, bands, etc. This  tool and all images used for the classification is made available publicly at  
\url{http://obas-matlas.u-strasbg.fr/}.
The participants were invited to fill a questionnaire about the presence, number and prominence of several types of features in or around the galaxies. Namely, we were interested in the following features (example images provided in Appendix~\ref{app:img} and on the MATLAS public web site):
\begin{itemize}
\item {\em The tidal features}, i.e. tails, streams and shells. These structures inform us about the gravitational interactions that the target galaxy had, or is having, with other galaxies. {\em Tails} refer to structures whose  material seems to come from the galaxy being classified. This means that it is either an ongoing interaction or remnant of a major merger and then the mass of the accreted galaxy constitutes a substantial fraction of the mass of target galaxy. {\em Streams} refer { to} structures apparently formed by a tidal disruption of a much smaller companion galaxy. {\em Shells}  are azimuthal arcs centered on the core of the host galaxy; they are characterized by sharp outer edges. Such features are known to form in minor to intermediate nearly radial mergers \citep{quinn83,bil15,pop18}. 
The measure of the frequency of each type of tidal features,  combined with knowledge of their specific life times and detectability as a function of time and viewing angles, estimated from simulations \citep{mancillas19}, can give us constrains on the past merging history of galaxies, and help to quantify the  fraction of stellar mass gathered by mergers of various types.
We should point out that this approach is valid for a given cosmological model. For instance,  assuming modified gravity, many tidal features are expected to have arisen from non-merging galaxy flybys \citep{bil18, bil19d}. 
\item {\em The shape of the outermost isophote} of the target galaxy. Strongly disturbed isophotes trace  recent interactions and are usually  associated with tidal features. Mildly irregular isophotes can be the last witnesses of old galaxy interactions when tidal features have already faded out. Alternatively, mild isophotal irregularities can indicate weak or just starting interactions. 
\item {\em { Peripheral disks}.} A characteristic of an ETG is an old red stellar population. Deep images however reveal that some ETGs are surrounded by faint { disks} of young blue stars, providing clues on the rejuvenation of ETGs. 
\item {\em The features induced by secular evolution (bars, rings, spiral arms).} Such features are usually located in the bright parts of the galaxy and thus shallow images are usually good enough for their detection. However they had not been previously identified in a systematic way in MATLAS. {Our deep data can reveal the exceptional cases of the secular features in the low-surface brightness regions.}
\item {\em Dust lanes and patches}. Such features can form spontaneously in the interstellar medium of ETGs but also trace past gas-rich merging events.
The MATLAS  deep images can  reveal dust regions in the outer stellar halo that are not detectable in shallower images. 
\item {\em Galactic cirrus}. Scattering the optical light of nearby stars, these dust clouds  trace the most diffuse interstellar medium of our own galaxy. They can complicate the detection and identification of the extragalactic structures under study, but are also interesting targets for detailed ISM studies at high angular resolution. 
\end{itemize}
{{In addition, two contributors logged  the presence of close-by {\em polluting halos}  that could have plausibly affected the detection or classification of the structures of interest. We considered two types of such halos: the instrumental ghosts surrounding  bright stars and caused by internal reflections in the camera (see Figures~\ref{fig:mosasymmetric}--\ref{fig:moscirrus} for examples), and the stellar halos of neighboring galaxies, whose isophotes overlap with those of the target.   
}}

{The answers of individual participants regarding a particular feature had to be converted to a single real number, the so-called rating of the feature. Initially, we converted the answers about the presence of the features in numbers, e.g. ``no'', ``likely'' and ``yes'' to 0, 1 and 2,  respectively. The final rating was chosen as the option that got most votes by the participants. If several answers got the same number of votes, the rating is the average of the most frequent answers. The rating of presence of a feature can range from 0, corresponding to absence of the feature, to 2, signifying the presence, with the exception of { peripheral disks}, whose ratings range between 0 and 1. 
{We assigned the galaxies a numerical rating of halos of 1 if the galaxies were affected by halos, or 0 otherwise.}
For some purposes, e.g., when we wanted to count how many galaxies have a certain type of feature, we had to round the rating to discrete categories. Then we speak about the ``rounded rating''.} }

\newcommand{\specialcell}[2][c]{%
  \begin{tabular}[#1]{@{}c@{}}#2\end{tabular}}
\begin{table}
\caption{Explanation of the classification codes.}
\label{tab:codes}
\begin{tabular}{l|l}
\hline\hline
Code & Meaning                                   \\
\hline
\multicolumn{2}{c}{Features}\\ \hline
symbol alone & the feature is present \\ 
symbol with ``?''    & the feature is likely / unsure \\
no symbol & the feature is missing \\
+s   & Streams \\
+r   & shells / Ripples                            \\
+t   & Tails                                     \\
+d   & { peripheral} star-forming Disk                  \\
+ah  & Asymmetric outer isophotes / stellar Halo \\
+ph  & \specialcell[t]{disturbed / Perturbed outer isophotes \\ \hspace{-8em} / stellar Halo}                 \\
+wl  & Weak dust Lanes                          \\
+pl  & strong / Prominent dust Lanes                         \\
+wb  & Weak Bars                                \\
+pb  & strong / Prominent Bars                              \\
\hline \multicolumn{2}{c}{Contaminants}\\ \hline
-h   & polluting Halos                               \\
-wc  & Weak Cirrus                              \\
-pc  & strong / Prominent Cirrus                        \\\hline
\end{tabular}
\end{table}

\begin{table*}
        \caption{Ratings and classification codes of the individual galaxies. The complete table is available in \app{ratings}.}
        \label{tab:ratings}
        \centering
        \begin{tabular}{lllllllllll}
                \hline\hline
                Name &  Classification code  &  Streams & Shells & Tails & Ext. SF & Out. isoph. & Dust & Bars & Halos & Cirrus \\\hline    
                
               IC\,0560      & +wb                  & 0.0      & 0.0      & 0.0      & 0.0      & 0.0      & 0.0      & 1.0      & 0.0      & 0.0     \\
IC\,0598      &                      & 0.0      & 0.0      & 0.0      & 0.0      & 0.0      & 0.0      & 0.0      & 0.0      & 0.0     \\
IC\,0676      & +pb                  & 0.0      & 0.0      & 0.0      & 0.0      & 0.0      & 0.0      & 2.0      & 0.0      & 0.0     \\
IC\,1024      & +s+ph+pl             & 2.0      & 0.0      & 0.0      & 0.0      & 2.0      & 2.0      & 0.0      & 0.0      & 0.0     \\
NGC\,0448     &                      & 0.0      & 0.0      & 0.0      & 0.0      & 0.0      & 0.0      & 0.0      & 0.0      & 0.0     \\
NGC\,0474     & +s+r+t+ph-h          & 2.0      & 2.0      & 2.0      & 0.0      & 2.0      & 0.0      & 0.0      & 1.0      & 0.0     \\
NGC\,0502     & +r+ah                & 0.0      & 2.0      & 0.0      & 0.0      & 1.0      & 0.0      & 0.0      & 0.0      & 0.0     \\
NGC\,0509     & +pb                  & 0.0      & 0.0      & 0.0      & 0.0      & 0.0      & 0.0      & 2.0      & 0.0      & 0.0     \\
NGC\,0516     & +wb                  & 0.0      & 0.0      & 0.0      & 0.0      & 0.0      & 0.0      & 1.0      & 0.0      & 0.0     \\
NGC\,0524     & +h?+wl-pc            & 0.0      & 0.5      & 0.0      & 0.0      & -1       & 1.0      & 0.0      & 0.0      & 2.0     \\

                \hline
        \end{tabular}
\end{table*}

\begin{table*}
	\caption{Census of the classified structures in the whole sample according to the rounded rating (in percent).}
	\label{tab:census}
	\centering
	\begin{tabular}{lllll}
		\hline\hline
Shells  & no: $84 \pm 7$   & likely: $5 \pm 2$   & yes: $12 \pm 3$   & unknown: $0 \pm 0$ \\ 
Streams  & no: $84 \pm 7$   & likely: $5 \pm 2$   & yes: $11 \pm 2$   & unknown: $0 \pm 0$ \\ 
Tails  & no: $87 \pm 7$   & likely: $3 \pm 1$   & yes: $10 \pm 2$   & unknown: $0 \pm 0$ \\ 
Outer isophotes  & regul.: $67 \pm 6$   & asym.: $12 \pm 3$   & disturb.: $20 \pm 3$   & unsure: $0.6 \pm 0.6$ \\ 
Tails or streams  & no: $76 \pm 7$   & likely: $5 \pm 2$   & yes: $19 \pm 3$   & unknown: $0 \pm 0$ \\ 
Shells or streams or tails  & no: $70 \pm 6$   & likely: $7 \pm 2$   & yes: $23 \pm 4$   & unknown: $0 \pm 0$ \\ 
Any tidal disturbance  & no: $59 \pm 6$   & likely: $13 \pm 3$   & yes: $28 \pm 4$   & unknown: $0 \pm 0$ \\ 
Bars  & no: $64 \pm 6$   & weak: $20 \pm 3$   & strong: $16 \pm 3$   & unsure: $0 \pm 0$ \\ 
Dust lanes  & no: $85 \pm 7$   & weak: $7 \pm 2$   & strong: $8 \pm 2$   & unsure: $0 \pm 0$ \\ 
Peripheral disks  & no: $96 \pm 7$ &    & yes: $4 \pm 1$   & unsure: $0 \pm 0$ \\ 
Halos  & no: $71 \pm 6$ &    & yes: $29 \pm 4$   & unsure: $0 \pm 0$ \\ 
Cirrus  & no: $83 \pm 7$   & weak: $7 \pm 2$   & strong: $10 \pm 2$   & unsure: $0 \pm 0$ \\ 
		\hline
	\end{tabular}
\end{table*}

\section{Results}
\label{sec:res}
\subsection{Final classification}
The ratings of the individual galaxies are presented in \tab{ratings}. Each galaxy was assigned a morphological code consistently with \citet{duc15} based on the rounded rating of the presence of the  classified features in the galaxy. The meaning of the code is explained in \tab{codes}. For example, the code {{``+s+d+ph+pl'' for IC\,1024 means: the galaxy contains streams (the ``+s'' symbol) but no shells, tails or bar (there are no symbols signifying these features); the outer isophote has a perturbed shape (the ``+ph'' symbol); the main body shows prominent dust lanes (the ``+pl'' symbol), and there is no cirrus in the field, or nearby halos  polluting the image}} (the corresponding symbols are missing). In Appendix~\ref{app:img} we present images of the galaxies that exhibit the most prominent examples of the classified structure types.


\subsection{Statistics of the structures}
Here we highlight some of the main results obtained from the statistical analysis of the visual classification. A detailed analysis, including the correlations between the frequency of the classified structures with the internal properties of the galaxies,  the comparisons with predictions from numerical simulations, and the conclusions about the past mass assembly of the target galaxies, will be presented in future papers of this series. 

The results of our census of the investigated features for our sample according to the rounded  rating are presented in \tab{census}. It shows the incidence the classified feature types in our sample along with the Poisson errors.

One can see that shells, streams and tails have about the same incidence appearing in about 10-15\% of galaxies. The more general indicator of tidal interactions, the irregularity of the outer isophotes, turns out to be more frequent. Isophotes are disturbed or asymmetric in about 30\% of the galaxies. Tidal structures however often appear together. Any of tails, streams and shells are at least likely in 30\% of galaxies. When we include galaxies with disturbed or asymmetric isophotes we detected signs of tidal interactions in about 40\% of galaxies. The distinction between steams and tails can be ambiguous. When these two categories are merged, we obtain that at least likely signs of them were detected in 20-25\% of galaxies.  The bars were rated at least as weak  in about 35\% of the sample.  Dust lanes were detected in about 15\% of galaxies. { Peripheral disks} are the least frequent type of  structures among those we investigated, appearing just in a few percent of all galaxies. {{Our census indicates that  40\% of the galaxies in our sample are not affected by the presence of polluting halos (from companions or bright stars).}} Galactic cirri, might have directly affected the classification of 10-20\% of galaxies.

\begin{figure*}
	\centering
	\includegraphics[width=17cm]{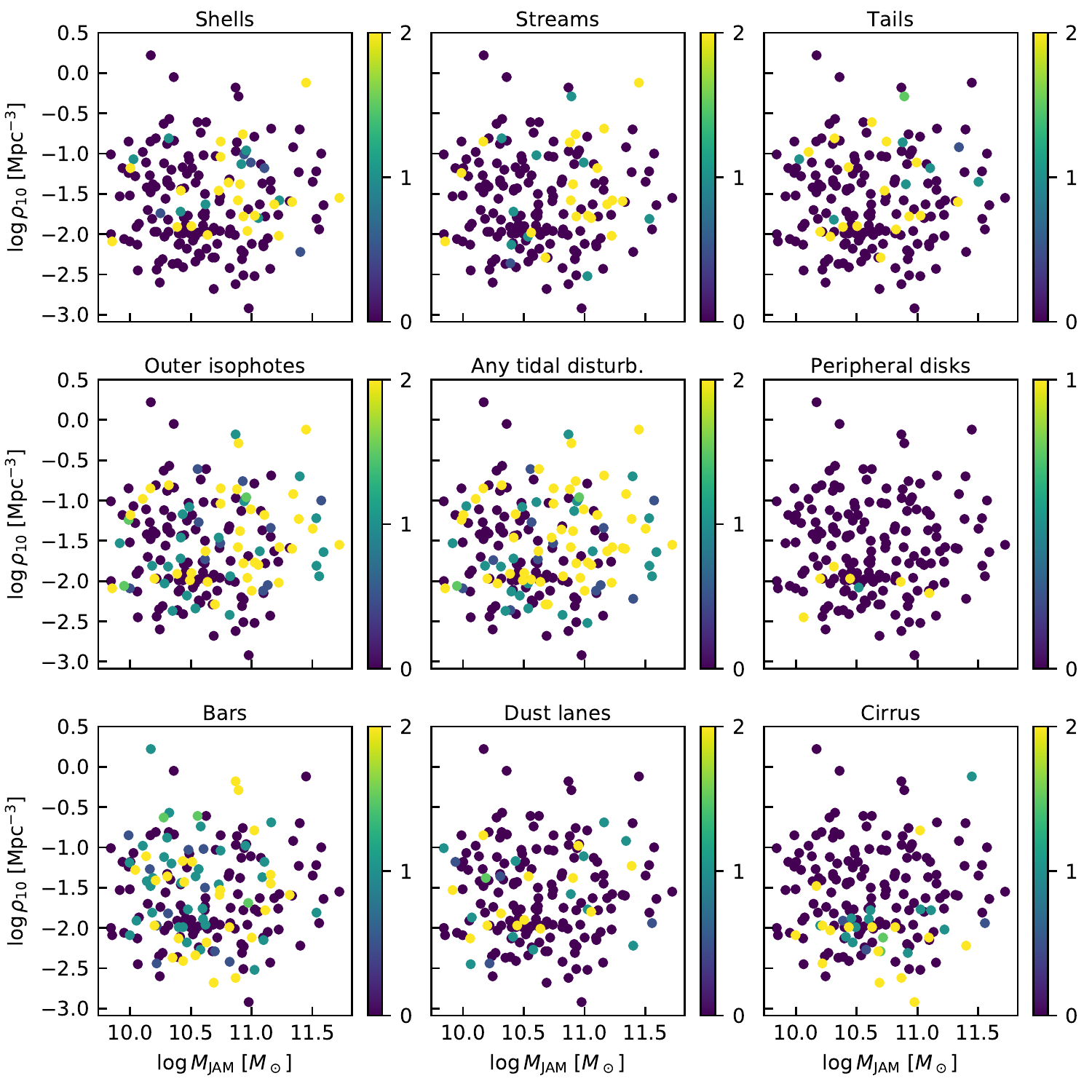}
        \caption{Occurrence of the investigated features depending on the  \mjam mass of the galaxy  and the density of its environment \rhoio. The color scale indicates the rating of the presence {or prominence} of the given feature in the particular galaxy.} 
        \label{fig:2d-all}
\end{figure*}

\begin{figure*}
	\centering
	\includegraphics[width=17cm]{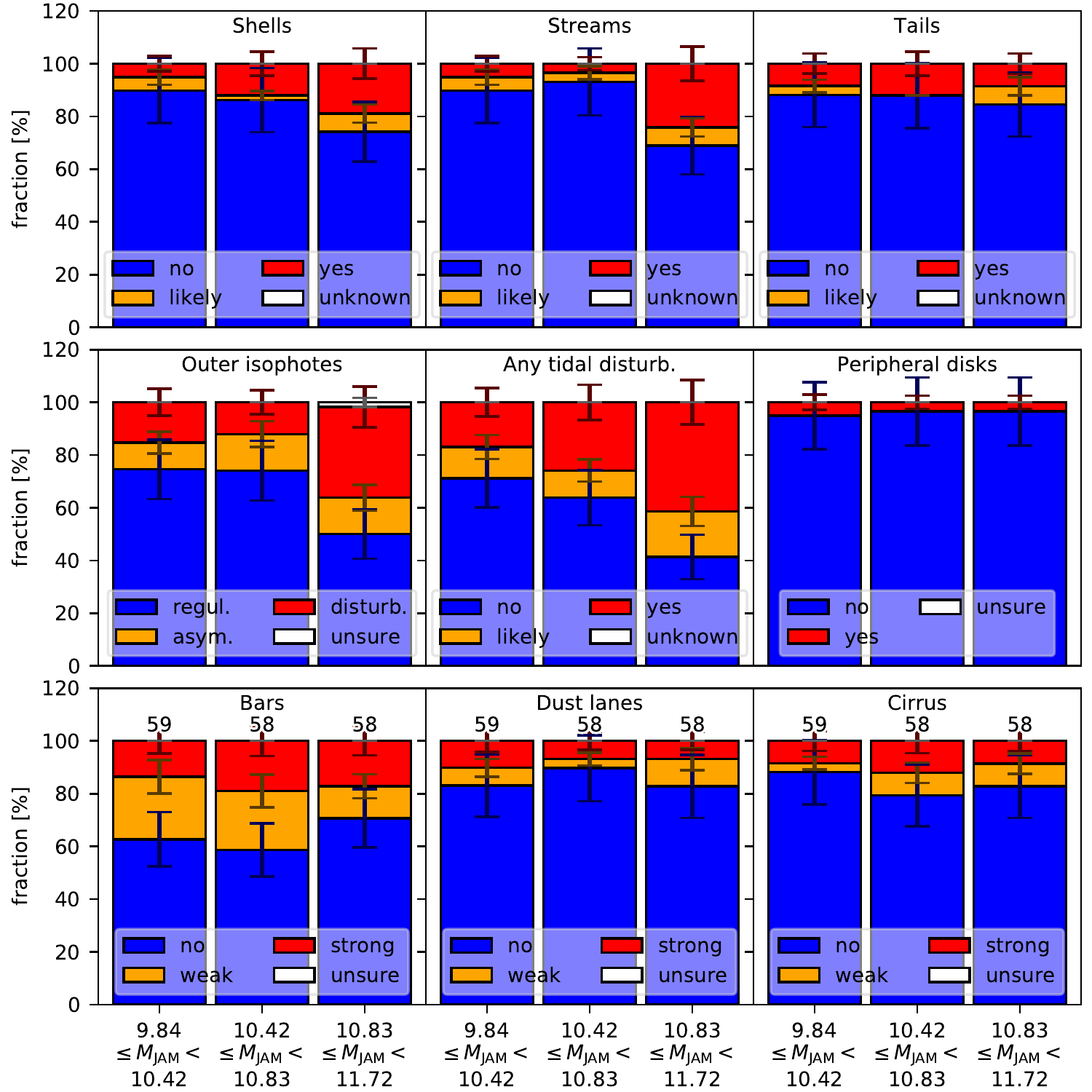}
	\caption{Fraction of the ratings of a given feature as a function of the \mjam mass of the galaxy (units $\log(\mjam/M_\odot)$). The rounded rating was used. The numbers over the bins in the bottom row of the panels denote the numbers of galaxies in the bins. They were equalized on purpose.}
	\label{fig:percmjam}
\end{figure*}

\begin{figure*}
	\centering
	\includegraphics[width=17cm]{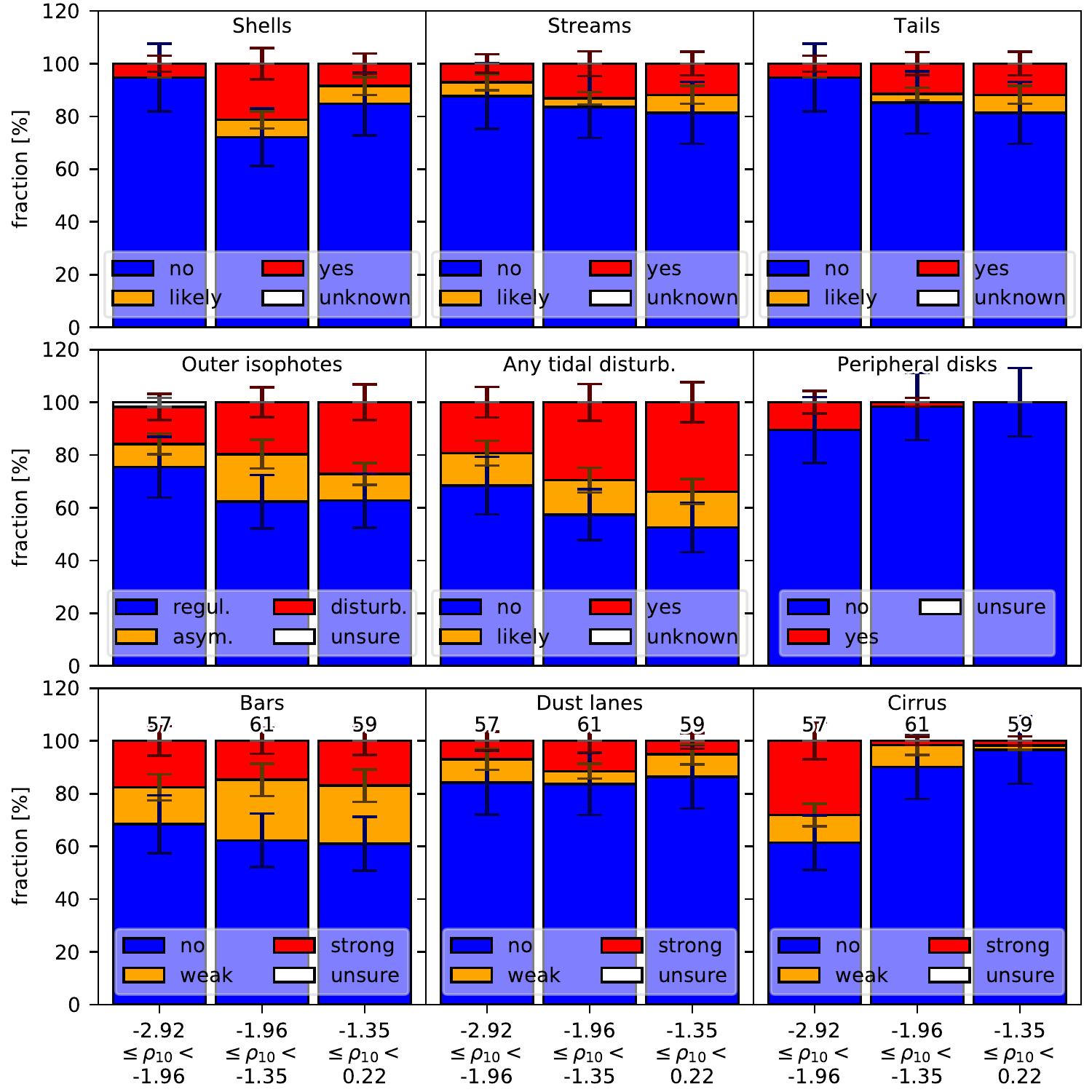}
		\caption{Fraction of the ratings of given feature as a function  environment density of galaxies \rhoio (units $\log(\rhoio/\mathrm{Mpc}^{-3})$). The rounded rating was used. The numbers over the bins in the bottom row of panels denote the numbers of galaxies in the bins. They were equalized on purpose. }
	\label{fig:percrho10}
\end{figure*}

We made also statistics of the number of tidal features per galaxy in \fig{freqhist}. The median number of the individual tidal-feature types is { 2.71 for shells, 1.0 for streams, and 1.25 for tails} where we have constrained ourselves just to the galaxies having the rating of the presence of the given feature  greater than zero  (galaxies in the zeroth bin contain, for example, one likely stream, i.e. the rating was 0.5). 

Theoretical arguments lead us to expect (see \sect{interp}) that the frequency of the investigated morphological  structures should depend on the mass of the galaxy and the density of its environment. We therefore evaluated how the frequency of fine structure  depends on the quantities \mjam and \rhoio.  The first of them, \mjam, is the dynamical mass obtained by Jeans Anisotropic Modeling \citep{cappellari08} within the sphere of a radius of one projected half-light radius as derived by \citet{cappellari13a} from the observed kinematic maps of the galaxies. \citet{cappellari13a} and \citet{poci17} calculated a median dark matter fraction of 13\% for the \atlas sample. Given the small fraction of dark matter needed, the mass \mjam is then a better estimator for the stellar mass than the estimates based on stellar population synthesis because of the uncertainties of stellar evolution, initial mass function and dust obscuration. The environment density \rhoio is defined as the mean density of galaxies inside a sphere centred on the galaxy and containing 10 nearest neighbours \citep{cappellari11b}. One might argue that mass and environment are closely related as more massive galaxies usually reside in denser environments but \fig{2d-all} demonstrates that $\mjam$ and $\rhoio$ are little correlated for our sample and that the incidence of the classified features depends on both quantities. After presenting the detected correlations in the rest of this section, we interpret them in \sect{interp}.

Figures~\ref{fig:percmjam} and~\ref{fig:percrho10} show how the frequency of the individual types of the classified structures depend on \mjam and \rhoio of the galaxy, respectively. The widths of the bins were set to contain equal number of galaxies. The two galaxies for which \mjam is not available, PGC\,058114 and PGC\,071531, were excluded.  

In addition, we made a statistical test of the significance of the correlations based on Spearman's rank coefficient, $r$, that is designed to be 1 (-1) for a strictly increasing (decreasing) sequence of data. The use of this coefficient is motivated by the visual impression of monotonic trends in Figs.~\ref{fig:percmjam} and~\ref{fig:percrho10}, e.g. the correlation between the occurrence of shells and the mass of the galaxy. The results are shown in \tab{corrsig}. Here we also give the corresponding $p$-value, i.e. the probability that the absolute value of $r$ is greater than the observed $r$ if there is actually no correlation between the occurrence of the feature with the property of the host galaxy. The values $r_\mathrm{all}$ and $p_\mathrm{all}$ were calculated for the whole sample. There are also trends in Figs.~\ref{fig:percmjam} and~\ref{fig:percrho10} that suggest a more complicated relationship, e.g. the peak in the occurrence of shells in the medium density bin or the vanishing of the correlation between the occurrence  of disturbed outer isophotes in the medium and low mass bins. We therefore divided the galaxy sample into two halves depending on whether the mass (environment density) of the galaxy was greater or smaller than the median mass (environment density) and applied the statistical test to each half; the median values { are 10.60 for} $\log \mjam/M_\odot$ and -1.64 for $\log \rhoio/(\rm{Mpc}^{-3})$. We list the results for the lower (higher) half of the sample in \tab{corrsig} as $r_\mathrm{1h}$ and $p_\mathrm{1h}$ ($r_\mathrm{2h}$ and $p_\mathrm{2h}$). The $p$-values that signify an inconsistency with no correlation at the 5\% confidence level are highlighted by boldface. One should look for a theoretical explanation of such correlations.

\begin{table*}
	\caption{Correlations of the occurrence the classified structures with the mass and environmental density of the host galaxy. The symbol $r$ stands for Spearman's rank coefficient and $p$ for the probability that $r$ is greater than the given value if there is actually no correlation. The subscript \textit{all} stands for the complete sample while \textit{1h} and \textit{2h} stand, respectively, for the correlation between the first and second half of the sample when sorted with respect to the mass or environment density. The $p$-values that are inconsistent with no correlation at the 5\% confidence level are highlighted by bold font.}
	\label{tab:corrsig}
	\centering
	\begin{tabular}{lllllll}
	\hline\hline
	Quantities & $r_\mathrm{all}$ & $p_\mathrm{all}$ & $r_\mathrm{1h}$ & $p_\mathrm{1h}$ & $r_\mathrm{2h}$ & $p_\mathrm{2h}$ \\
		\hline
Shells -- $\log M_\mathrm{JAM}$ & 0.22 & $\mathbf{0.0031}$ & -0.053 & 0.63 & 0.11 & 0.29 \\ 
Streams -- $\log M_\mathrm{JAM}$ & 0.21 & $\mathbf{0.0049}$ & 0.017 & 0.87 & 0.20 & 0.064 \\ 
Tails -- $\log M_\mathrm{JAM}$ & 0.074 & 0.33 & -0.012 & 0.91 & -0.017 & 0.88 \\ 
Out. iso. -- $\log M_\mathrm{JAM}$ & 0.27 & $\mathbf{0.00028}$ & -0.035 & 0.74 & 0.32 & $\mathbf{0.0023}$ \\ 
Any tidal disturb. -- $\log M_\mathrm{JAM}$ & 0.30 & $\mathbf{0.000041}$ & 0.026 & 0.81 & 0.26 & $\mathbf{0.016}$ \\ 
Peripheral disks -- $\log M_\mathrm{JAM}$ & -0.064 & 0.40 & 0.028 & 0.80 & 0.018 & 0.87 \\ 
Secular f. -- $\log M_\mathrm{JAM}$ & -0.092 & 0.23 & 0.12 & 0.26 & -0.14 & 0.21 \\ 
Dust -- $\log M_\mathrm{JAM}$ & -0.063 & 0.40 & -0.10 & 0.35 & 0.20 & 0.067 \\ 
Cirrus -- $\log M_\mathrm{JAM}$ & 0.073 & 0.34 & 0.070 & 0.52 & -0.073 & 0.50 \\ 
Shells -- $\log \rho_{10}$ & 0.14 & 0.065 & 0.27 & $\mathbf{0.013}$ & -0.14 & 0.19 \\ 
Streams -- $\log \rho_{10}$ & 0.080 & 0.29 & 0.073 & 0.50 & 0.066 & 0.54 \\ 
Tails -- $\log \rho_{10}$ & 0.16 & $\mathbf{0.035}$ & 0.20 & 0.058 & 0.14 & 0.20 \\ 
Out. iso. -- $\log \rho_{10}$ & 0.13 & 0.090 & 0.18 & 0.091 & 0.013 & 0.91 \\ 
Any tidal disturb. -- $\log \rho_{10}$ & 0.14 & 0.064 & 0.099 & 0.36 & -0.011 & 0.92 \\ 
Peripheral disks -- $\log \rho_{10}$ & -0.18 & $\mathbf{0.014}$ & -0.021 & 0.85 & nan & nan \\ 
Secular f. -- $\log \rho_{10}$ & 0.067 & 0.37 & -0.14 & 0.19 & 0.010 & 0.93 \\ 
Dust -- $\log \rho_{10}$ & -0.015 & 0.85 & 0.027 & 0.81 & -0.024 & 0.82 \\ 
Cirrus -- $\log \rho_{10}$ & -0.37 & $\mathbf{0.00000035}$ & -0.18 & 0.10 & 0.10 & 0.33 \\ 
		\hline
	\end{tabular}
\end{table*}

\begin{table*}
	\caption{Census of the classified structures in the whole sample according to the rounded rating (in percent) but only for the galaxies whose \mjam mass exceeds $10^{11}$\,M$_\odot$.}
	\label{tab:census_upperM}
	\centering
	\begin{tabular}{lllll}
		\hline\hline
Shells  & no: $80 \pm 10$   & likely: $6 \pm 4$   & yes: $17 \pm 7$   & unknown: $0 \pm 0$ \\ 
Streams  & no: $70 \pm 10$   & likely: $6 \pm 4$   & yes: $26 \pm 9$   & unknown: $0 \pm 0$ \\ 
Tails  & no: $90 \pm 20$   & likely: $6 \pm 4$   & yes: $6 \pm 4$   & unknown: $0 \pm 0$ \\ 
Outer isophotes  & regul.: $50 \pm 10$   & asym.: $17 \pm 7$   & disturb.: $30 \pm 10$   & unsure: $3 \pm 3$ \\ 
Tails or streams  & no: $60 \pm 10$   & likely: $9 \pm 5$   & yes: $29 \pm 9$   & unknown: $0 \pm 0$ \\ 
Shells or streams or tails  & no: $60 \pm 10$   & likely: $9 \pm 5$   & yes: $31 \pm 9$   & unknown: $0 \pm 0$ \\ 
Any tidal disturbance  & no: $40 \pm 10$   & likely: $20 \pm 8$   & yes: $40 \pm 10$   & unknown: $0 \pm 0$ \\ 
Bars  & no: $70 \pm 10$   & weak: $17 \pm 7$   & strong: $17 \pm 7$   & unsure: $0 \pm 0$ \\ 
Dust lanes  & no: $80 \pm 20$   & weak: $11 \pm 6$   & strong: $6 \pm 4$   & unsure: $0 \pm 0$ \\ 
Peripheral disks  & no: $100 \pm 20$ &    & yes: $3 \pm 3$   & unsure: $0 \pm 0$ \\ 
Cirrus  & no: $80 \pm 20$   & weak: $11 \pm 6$   & strong: $9 \pm 5$   & unsure: $0 \pm 0$ \\ 
		\hline
	\end{tabular}
\end{table*}

\begin{table*}
	\caption{Census of the classified structures in the whole sample according to the rounded rating (in percent) but only for the galaxies whose environmental density is below $\log_{10} \rhoio/(1\,\mathrm{Mpc}^{-3}) = -2$.}
	\label{tab:census_lowerrho}
	\centering
	\begin{tabular}{lllll}
		\hline\hline
Shells  & no: $90 \pm 10$   & likely: $0 \pm 0$   & yes: $7 \pm 4$   & unknown: $0 \pm 0$ \\ 
Streams  & no: $90 \pm 10$   & likely: $7 \pm 4$   & yes: $7 \pm 4$   & unknown: $0 \pm 0$ \\ 
Tails  & no: $100 \pm 10$   & likely: $0 \pm 0$   & yes: $4 \pm 3$   & unknown: $0 \pm 0$ \\ 
Outer isophotes  & regul.: $70 \pm 10$   & asym.: $11 \pm 5$   & disturb.: $13 \pm 5$   & unsure: $2 \pm 2$ \\ 
Tails or streams  & no: $80 \pm 10$   & likely: $7 \pm 4$   & yes: $11 \pm 5$   & unknown: $0 \pm 0$ \\ 
Shells or streams or tails  & no: $80 \pm 10$   & likely: $7 \pm 4$   & yes: $13 \pm 5$   & unknown: $0 \pm 0$ \\ 
Any tidal disturbance  & no: $70 \pm 10$   & likely: $16 \pm 6$   & yes: $18 \pm 6$   & unknown: $0 \pm 0$ \\ 
Bars  & no: $70 \pm 10$   & weak: $13 \pm 5$   & strong: $18 \pm 6$   & unsure: $0 \pm 0$ \\ 
Dust lanes  & no: $80 \pm 10$   & weak: $11 \pm 5$   & strong: $4 \pm 3$   & unsure: $0 \pm 0$ \\ 
Peripheral disks  & no: $90 \pm 10$ &    & yes: $9 \pm 4$   & unsure: $0 \pm 0$ \\ 
Cirrus  & no: $60 \pm 10$   & weak: $9 \pm 4$   & strong: $27 \pm 8$   & unsure: $0 \pm 0$ \\ 
\hline
	\end{tabular}
\end{table*}

In \fig{percmjam}  we can see that tidal features are more frequent in more massive galaxies. The correlations for streams, shells and perturbed isophotes are statistically significant while that for the tidal tails is not as we can learn from \tab{corrsig}. The field  ``Any tidal disturbance'' indicates whether the galaxy has either shells, streams, tails or irregular isophotes; more precisely this quantity is defined as the maximum of the ratings of these morphological features. There is a visual appearance of an abrupt boost of the incidence of streams, disturbed isophotes {{or tidal disturbances in general}} at high masses. { The confidence of this feature misses our adopted confidence threshold of 5\%.}
There are also no significant correlations of the incidence of bars, dust lanes, and, as expected, foreground cirrus with the mass of the galaxy.

Table~\ref{tab:census_upperM} provides the census of the fine structures for galaxies over the mass of $10^{11}$\,M$_\odot$. { There are 35 such galaxies.}  Comparison to the census in the complete sample in \tab{census} yields that the incidence
of shells, streams and disturbed isophotes increases, for these massive galaxies, { by factors of  $1.4\pm0.5$,  $2.0\pm0.6$ and  $1.6\pm0.4$, respectively.  }

In \fig{percrho10} we present the correlations of the incidence of our structures of interest with the environment density of the target galaxy, probed by the \rhoio parameter. The feature that correlates the strongest with the galaxy environment is surprisingly the presence of cirrus in our own Galaxy. The lower galaxy density, the more probable the cirrus occurrence is. We discuss the possible explanations of this unexpected effect in \sect{disc}. There is also a statistically significant correlation of the environment density with the incidence of tails. { On the contrary, the incidence of peripheral disks decreases significantly with increasing the density of the environment.  Apart from this, the shell incidence increases significantly with \rhoio in the low-density half of the sample. } The decrease of the incidence of shells between the medium and high density bins is not confirmed by the test. Neither are the apparent correlations between the incidence of streams and disturbed isophotes with \rhoio. {{The correlation of \rhoio with the union of all tidal disturbances closely misses our significance threshold.}}

Table~\ref{tab:census_lowerrho} provides the census of the fine structures for the 45 galaxies residing in environments whose densities are below $10^{-2}$\,Mpc$^{-3}$. By comparing to \tab{census}, one can find that the incidence
of shells and tails is, compared to the whole sample, lower by a factor of  $0.4\pm0.2$ and  $0.3\pm0.2$, respectively.

Dust lanes and  bars  do not show any correlation with environment density of the host.  We recall here that our sample does not include cluster galaxies, i.e. the galaxies in the densest environments.

Besides mass and environment, the incidence of the features studied here might correlate with other properties of the galaxy, such as the specific angular momentum, gas content, presence or absence of a core, etc. These will be studied in a future paper.

\section{Discussion}
\label{sec:disc}

\begin{table*}
	\caption{Census of the classified structures according to the rounded rating (in percent). Only galaxies without strong pollutants were considered (rating of {polluting halos} $=1$, rating of  cirrus $<1.5$).}
	\label{tab:expcensusexclude}
	\centering
	\begin{tabular}{lllll}
		\hline\hline
Shells  & no: $89 \pm 9$   & likely: $2 \pm 1$   & yes: $9 \pm 3$   & unknown: $0 \pm 0$ \\ 
Streams  & no: $87 \pm 9$   & likely: $3 \pm 2$   & yes: $11 \pm 3$   & unknown: $0 \pm 0$ \\ 
Tails  & no: $95 \pm 9$   & likely: $3 \pm 2$   & yes: $3 \pm 2$   & unknown: $0 \pm 0$ \\ 
Outer isophotes  & regul.: $73 \pm 8$   & asym.: $12 \pm 3$   & disturb.: $15 \pm 4$   & unsure: $0 \pm 0$ \\ 
Tails or streams  & no: $76 \pm 7$   & likely: $5 \pm 2$   & yes: $19 \pm 3$   & unknown: $0 \pm 0$ \\ 
Shells or streams or tails  & no: $70 \pm 6$   & likely: $7 \pm 2$   & yes: $23 \pm 4$   & unknown: $0 \pm 0$ \\ 
Any tidal disturbance  & no: $59 \pm 6$   & likely: $13 \pm 3$   & yes: $28 \pm 4$   & unknown: $0 \pm 0$ \\ 
Bars  & no: $59 \pm 7$   & weak: $24 \pm 5$   & strong: $17 \pm 4$   & unsure: $0 \pm 0$ \\ 
Dust lanes  & no: $82 \pm 9$   & weak: $7 \pm 3$   & strong: $11 \pm 3$   & unsure: $0 \pm 0$ \\ 
Peripheral disks  & no: $96 \pm 9$ &    & yes: $4 \pm 2$   & unsure: $0 \pm 0$ \\  
		\hline
	\end{tabular}
\end{table*}

\subsection{Discussion of our method}
\label{sec:discmeth}
As  already pointed out in the Introduction, the visual classification of the faint structure is necessarily subjective. Different people can classify a given feature as a tail or stream or these could be confused with Galactic cirrus. 
Nevertheless,  before objective automatic algorithms for structure detection and classification improve, visual identification remains the best option.

Our adopted rating procedure has several desirable properties. Since it includes voting, it eliminates the mistakes of individuals. We based our results on the votes of the participants who inspected at least two-thirds of the galaxies. Such classifiers had already a substantial experience with controlling the navigation tool, visualizing the faint structures, and distinguishing tails and streams, for example. { We made a few tests to verify this expectation, see \app{experience}.} As the disadvantage, the voting method eliminates correct identifications of indistinct structures by sensitive individuals.

We explored the influence of the image pollutants, the {{polluting halos}} and cirri, in \tab{expcensusexclude}. It shows the census of our structures of interest but calculated only considering the galaxies whose images were not polluted much, namely the rating of the {{polluting halos}} was 0 and the rating of cirrus was $<1.5$. Comparing \tab{expcensusexclude} to \tab{census} one can see that there is no significant difference. {{This suggests that these pollutants do not affect our classification substantially.}}  

At the time of the census,  model subtracted images were not available for the majority of our galaxies. It is very likely that we missed some tidal features because of this. Tidal features can be hard to detect in the central parts where the luminosity of the underlying galaxy has a steep gradient. For this reason,  the number of tidal features is very likely underestimated. On the other hand, our deep images enabled us to detect well the structures in the outer parts of the galaxies that could be overlooked in the standard shallow images.

\subsection{Comparison to literature}
\label{sec:lit}
In this section we compare our results to other works. 

\subsubsection{Tidal structures}
 Table~1 of \citet{atkinson13} provides a useful compilation of tidal feature occurrence from 12 different works. The fraction of galaxies with tidal features varies a lot, from 3 to over 70\%. This is probably not only a result of different instruments, selection criteria, image depths and image processing techniques, but also of different criteria on the prominence of the feature. For example, \citet{atkinson13} give for their own results several degrees of confidence that a galaxy contains a tidal feature. Constraining themselves on the red galaxies in the highest confidence category, they found tidal features in 15\% of galaxies, while counting in all red galaxies with any signs of tidal features, they arrived to 41\%. The work of \citet{atkinson13} is  similar to  ours since they used a visual identification on images obtained with the same instrument. They however processed and stacked differently, with a method not optimized for LSB studies. 
  Their morphological categories differ from  ours. One has to compare our category of shells with the union of their categories of shells, fans and diffuse structures; some of the features classified as the diffuse structures by \citet{atkinson13} would however be classified by us as galaxies with disturbed outer isophotes. Constraining ourselves just to their red galaxies and their two highest-confidence detection levels, their sample contains  12\% of shell galaxies. This is consistent with the measured likely or secure  shell occurrence   of $17\pm5\%$ in the MATLAS sample. This is more than \citet{MC83} who detected shells in around 10\% of their galaxies, probably because their limiting surface brightness  was shallower by about 2 mag/arcsec$^{-2}$. In contradiction with us, \citet{MC83} found that the frequency of shells decreases with an increasing environment density. We detected a hint of peak of shell occurrence in the medium density bin but the decrease toward the high densities is not statistically significant. This might be explained by the fact that our sample does not include galaxies in clusters.

 It seems the most relevant to compare the union of our categories of streams and tails with the union of the structures called streams, linear and arms in the red galaxies of \citet{atkinson13}.  The comparison yields $24\pm7\%$ of these types of tidal features in our sample versus $14\pm1\%$ in \citet{atkinson13}.  
Besides the difference in the data reduction technique between the two surveys, 
 one of the main reasons of this inconsistency is probably the larger average distance to their galaxies that { made the detectability of these thin and faint structures harder}. Indeed, the median redshift of their red galaxies is 1.4, i.e. around 500\,Mpc, while our galaxies lie in the median distance of only 27.2\,Mpc. Moreover, it seems from the example images provided by \citet{atkinson13} that some of their structures classified as ``diffuse'' would likely be classified as tails by us.
 Shell detectability is probably not affected that much by the distance  perhaps because shells are more sharp-edged and are thus more easy to detect.

We detected a tidal disturbance of any type in $41\pm7\%$ of galaxies, counting in also the likely detections. Red galaxies in the sample of \citet{atkinson13} have some form of a tidal disturbance in $22\pm2\%$. We attribute this to the larger distances to the galaxies in \citet{atkinson13} since a lower angular size and a lower amount of captured light can preclude the visibility of faint streams and asymmetric isophotes.

\citet{atkinson13} found that the occurrence of tidal features in red galaxies increases with the mass of the galaxy. We detected this just for streams, shells and disturbed isophotes (not tails). They did not investigate the correlation with environment density.

\citet{pop18} presented a census of shells in galaxies with masses over $10^{11}$\,M$_\odot$ in a cosmological hydrodynamical simulation, finding the shell incidence of 20-30\%. This agrees well with{ our finding of $23\pm8\%$ of at least likely detections in our sample} when the same mass cut is applied. They however did not apply any limits on the surface brightness.

\subsubsection{Secular features}
We detected strong or weak secular features (bars, spiral arms, or rings but mostly bars) in $36\pm6\%$ of galaxies. For comparison \citet{krajnovic11} detected, in the very same sample (i.e. after excluding the \atlas galaxies that do not belong to the MATLAS sample), $28\pm4\%$ galaxies with bars or rings but in images taken with 2.5\,m optical telescopes. 

 \citet{laurikainen13} investigated the occurrence of bars in lenticular galaxies in a magnitude limited sample using near-infrared ground-based images taken by 3-4\,m class telescopes. They found that the bar occurrence depends on the Hubble stage number of the galaxy increasing from $\sim 35\%$ for the -3 type to $\sim 75\%$ for the -1 type. We calculated from their Table~1 that they found $57\pm6\%$ of barred galaxies among the morphological types from -3 to 0. To make a fair comparison, we obviously had to restrict ourselves to lenticular galaxies in our sample. Additionally, we wanted to minimize the effect of the different fraction of the morphological sub-classes of lenticular galaxies in the two samples.  This led us to consider the Hubble stage number for our galaxies given in \citet{cappellari11a} and divide the galaxies into several bins centered on the integer stage numbers and having widths of one stage number. We multiplied the numbers of both barred and nonbarred galaxies in each bin by a constant so that the total number of galaxies in the bin was the same as in the corresponding bin of \citet{laurikainen13}. The  $57\pm6\%$ of barred galaxies in the sample of \citet{laurikainen13} should be compared to $44\pm7\%$ of barred galaxies in our sample. Our results are therefore consistent with  \citet{laurikainen13}.  We counted as barred those of our galaxies having the rounded rating of at least 1 (i.e., loosely speaking, at least weak bars). The error for our sample was estimated in a Monte-Carlo way.

As for the correlations of bar occurrence with mass and environment density, \citet{wilman12} divide galaxies to central and non-central with respect to their group. They found evidence that the bar incidence depends on the stellar mass of the galaxy only for the low-mass central galaxies. For them, the bar incidence is enhanced with respect to the level defined by the non-central galaxies. \citet{barway11} demonstrated that bar occurrence in lenticulars depends on mass and environment. They found that the bar fraction decreases with luminosity of the galaxy and that the bar fraction increases with the environment density. The dependence on the environment density is stronger for faint galaxies. Interestingly, bar occurrence does not depend on environment density for a general population of disk galaxies, i.e. consisting of both lenticular and spiral galaxies \citep{aguerri09,martinez11,lin14}.

In order to make a comparison with these older works, we had to restrict ourselves only to the lenticular galaxies in our sample, i.e. to morphological types between -3.5 and 0.5. { For such galaxies, the sample as a whole does not correlate significantly with \mjam.}The incidence of bars neither correlates with \rhoio, neither for the whole sample, nor with the low- or high-density halves. There is just a hint of correlation for the low-density half of the lenticular sample (density below $\log_{10} \rhoio/(\mathrm{Mpc}^{-3}) = -1.63$) -- bar occurrence decreases with environment (Spearman coefficient of -0.2) at the 8\% confidence level. We have to remind here again that our survey, unlike the previous works, excludes galaxy clusters, i.e. the densest environments. Similarly, the correlation of bar occurrence with the \mjam mass is not significant for the total sample of our lenticulars. We however detect, contrarily to the literature results, that the bar occurrence increases with galaxy mass for the low-mass half of the sample of lenticulars { (i.e., mass below $\log_{10} \mjam/M_\odot <10.56$)  at a statistically significant confidence level of 4\%.}

\subsubsection{Dust}
We can learn about the history of dust detection in elliptical galaxies from the review by \citet{kormendy89}. The authors say that many dust clouds in ellipticals are small and their detection depends critically on the resolving power of the instrument and seeing. Apart from this, the detection also requires dividing the image by a smooth model of the galaxy, which we did not do. This is perhaps the reason why we detected traces of dust only in $15\pm4\%$ of our galaxies. \citet{kormendy89} state a dust occurrence between 20 and 40\%.

 Older works report an absence of correlation between dust mass and stellar mass or luminosity (e.g., \citealp{smith12,hirashita15,kokusho19}). In order to verify this in our sample, { we restricted our analysis to elliptical galaxies, i.e. the} morphological types below -3.5. There is indeed no significant correlation with \mjam (the $p$-value of the Spearman coefficient is over the 5\% threshold, even for the low- or high-mass halves).  

\subsubsection{{ Peripheral disks}}
We are not aware of any other precise statistics of the occurrence of { peripheral star-forming disks in ETGs. A few  ETGs in the MATLAS sample with evidence of peripheral star-forming disks and associated extended disk of atomic hydrogen have been studied in \citet{Yildiz17}. Galaxies with  extended star formation  (which is also visible in the UV survey by GALEX) actually resemble massive LSB galaxies such as Malin\,1 \citep{Galaz15,Boissier16}.  The latter consist of a faint disk, which is actually more extended than in our ETGs, and a prominent bulge.    
We note that our nearby ETGs with { peripheral disks} might be local analogs of galaxies at high redshifts: \citet{sachdeva19} suggested that disks frequently form around pre-existing bulges at the redshift of 2. Some of the  peripheral disks might also possibly be analogs of polar rings appearing however in the equatorial plane of the galaxy}.

\subsection{Interpreting the results}
\label{sec:interp}
Let us draw preliminary conclusions from our results, although an exhaustive analysis is beyond the scope of this paper. The incidences of structures  have to be compared with the  numbers predicted by theoretical models of galaxy formation. Here we thus focus just on tentative  qualitative theoretical explanations for the observed trends of the incidences with the mass and environment  of the target galaxy.

We detected statistically significant increase of the incidence of shells, streams and disturbed outer isophotes with the mass of the galaxy.  We can think of several reasons for this. More massive galaxies have stronger tidal forces that can disrupt more massive neighbors and in a greater distance. The debris of bigger galaxies are also probably observable for a longer time. A more massive galaxy can attract its neighbors from larger distances and disrupt them afterward. Apart from this, more massive galaxies usually have a larger number of satellites and reside in denser environments such that there are more objects available for disruption. In a hierarchical model of galaxy formation, a higher abundance of tidal features in massive galaxies is expected since more mass was accreted by these galaxies.

\begin{figure}
        \resizebox{\hsize}{!}{\includegraphics{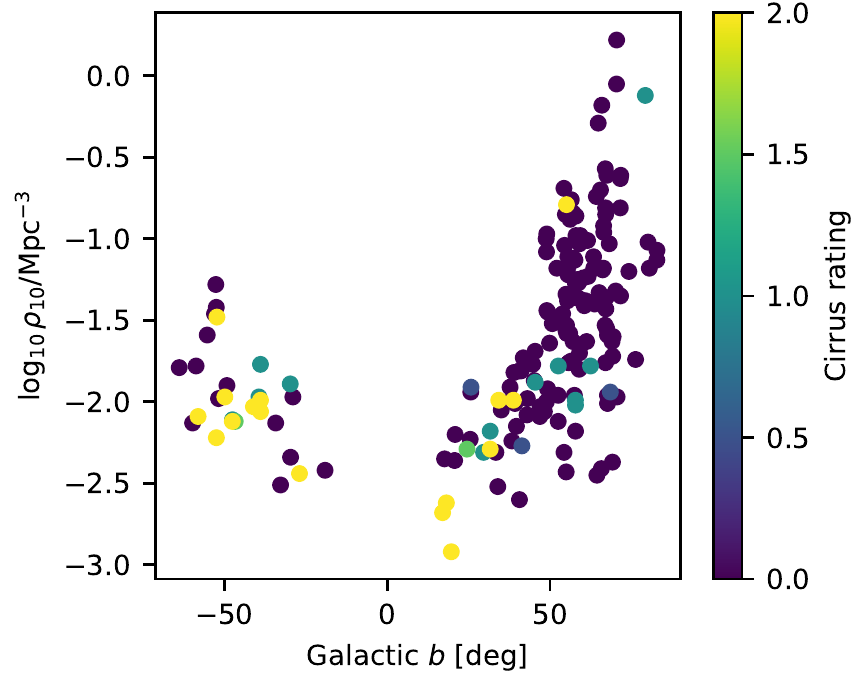}}
        \caption{Environment densities of external galaxies depend on Galactic latitude.} 
        \label{fig:galb}
\end{figure}

\begin{figure}
        \resizebox{\hsize}{!}{\includegraphics{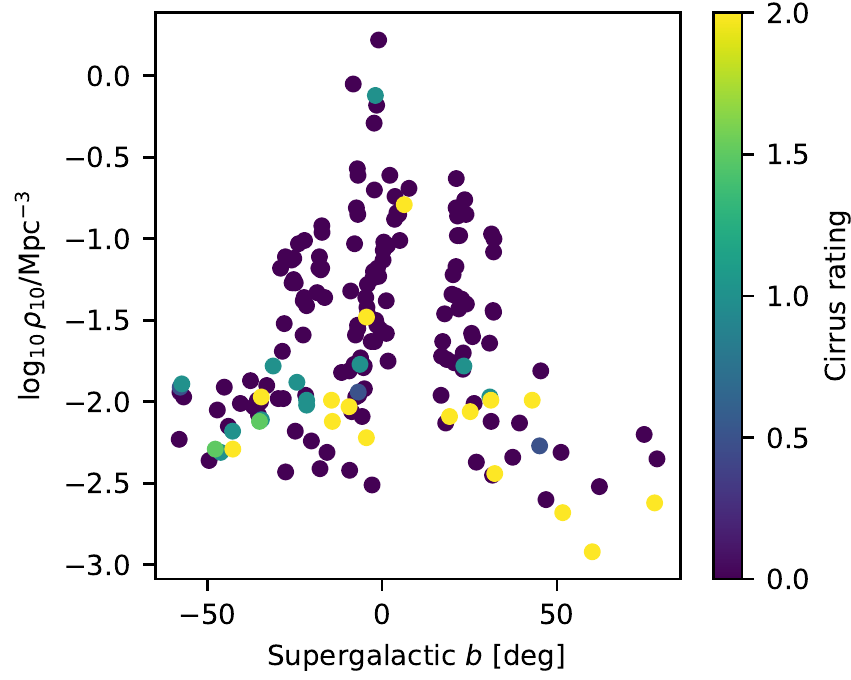}}
        \caption{Environment densities of external galaxies depend on Supergalactic latitude.} 
        \label{fig:supergalb}
\end{figure}

We found hints of correlations of the incidence of all types of tidal features with environment density. However, only two of them are statistically significant: the one for the incidence of tails and that for the incidence of shells in the low-density half of the sample. Indeed, we expect more frequent galaxy encounters in denser environments. The peak in the abundance of shells in the medium density bin is not statistically significant with the current data but there is a theoretical motivation for it. It is the result of two competing factors. On one hand, galaxy encounters are less likely in sparse environments. On the other hand, theoretical arguments suggest that shells form more likely if the accreted galaxy, that subsequently turns into the shells, is a disk galaxy \citep{HQ88} and it is known that disk galaxies are less frequent in high density environments. Moreover, stars released from the accreted galaxy can be so fast in the high-velocity encounters in the dense environments that they exceed the escape velocity, or dynamical friction during galaxy high-velocity encounters is ineffective. 
Many of the tails extend from galaxies that appear being disrupted by a more massive neighbor.



{ We found an anticorrelation of the incidence of peripheral star-forming disks with the environment density.} It is in line with the well-known relation between the gas content of disk galaxies and environment -- gas rich spiral galaxies are frequent in sparser  environments while gas poor lenticulars are found rather in denser environments. One reason for this is ram pressure stripping of gas as the galaxy moves quickly through the intercluster/intergroup medium. If the ram pressure is not strong enough to remove the gas from the galaxy, star formation is quenched by the starvation mechanism when the surrounding hot intergroup medium prevents the accretion of the cold intergalactic medium that is available in the low density environment to feed the star formation in the galaxy.  Star formation can also be quenched because of the gas is shocked and heated by galaxy interactions \citep{bitsakis16,ardila18,bitsakis19}.

Most surprisingly, the strongest correlation we found is that between the occurrence of Galactic cirrus and the environment of the background galaxy -- the stronger cirrus, the lower environment density.  We propose two possible explanations.
One is based on the hypothesis that the environment density was underestimated at low galactic latitudes because of dust obscuration, as \fig{galb} suggests. The cirri are located close to the Galactic plane as well. 
 There are however two arguments against this hypothesis. First, the incidence of Galactic cirrus does not correlate that well with a small-scale environment density $\Sigma_3$, defined as the mean projected density in a cylinder containing three nearest neighbors, a  quantity from \citet{cappellari11b}. The $p$-value of the correlation with $\Sigma_3$ is 0.4\%, while that of the correlation with $\rho_{10}$ is $10^4$ times lower. Besides, the environment density \rhoio was determined from the $K$-band near-infrared images from the Two-Micron All-Sky Survey \citep{2mass}. Even the $B$-band extinction for our galaxies listed in \citet{cappellari11a} is not high enough to explain the drop of environment density, when the Schechter function is taken into account, and this extinction is typically ten times greater than in the $K$-band.

We find it most probable that the trend  is an indirect consequence of the orientation of the Milky Way disk with respect to the local structure of the cosmic web. It is well known that the Milky Way disk is nearly perpendicular to the Local Sheet, which defines the equatorial plane of the supergalactic coordinate system. The environment density indeed increases with the supergalactic latitude of the target galaxy, see \fig{supergalb}.

\section{Future prospects}
\label{sec:future}
The visual classification method adopted in this paper allows us to distinguish the various types of tidal features, which give constraints on the mass assembly of the galaxy (e.g., minor/major or dry/wet mergers). Tidal features bring even information about the collision date because  diffuse and sharp-edged tidal features have different survival times  at out observation depth \citep{mancillas19}.

Obviously the main drawback of the visual method is its intrinsic subjectivity (see the discussion above) and its unfeasibility on very large samples. The forthcoming deep and wide-field surveys (LSST, CFIS, Euclid) create pressure to develop automated algorithms able to detect and classify tidal features in an  enormous number of galaxies (see e.g., \citealp{laine18}). The use of automatic methods, for instance based on a deep learning (DL) approach, will become inevitable. However such algorithms will not necessarily  allow us to overcome the subjective nature of the visual inspection, as DL requires to compile a training set, also relying on a classification by eye, unless numerical simulations are used. One major difficulty when trying to distinguish with DL different types of tidal features with subtle differences in their morphology, such as streams and tails, is the currently small sample of training images that had previously been annotated visually. Under such conditions, further developments are needed to guide the neural networks in the learning process. 

Besides, one wishes to go beyond the qualitative methods which consists of just counting the number of fine features or estimating their strength. Getting quantitative requires doing the proper photometry of the collisional debris. This remains a challenge given their low surface brightness nature. A number of techniques are being developed to automatically trace faint features and obtain precise measurements of the sky background level \citep{Akhlaghi15}. However the level of fine tuning they require still prevents from applying them to large surveys, including MATLAS.

\section{Summary}
\label{sec:sum}
In this paper we presented the first results from the MATLAS deep-optical-imaging volume-limited survey of 177 nearby ETGs that were drawn from the \atlas sample \citep{cappellari11a} of well-explored galaxies. Our goal was to survey the incidence of the following structures in the galaxies: tidal streams, tails, and shells, the irregularity of outermost isophotes, bars in the centers of galaxies, dust lanes, presence of star formation at the outskirts of the galaxy, and the presence of close or interacting companions. Additionally, we investigated the incidence of two frequent pollutants of the images -- {{polluting halos}} and galactic cirri. The detection and classification was performed visually by a group of researchers who had a substantial previous experience with the investigations of ETGs and interacting galaxies. 
The results presented here are based on votes of six individuals who inspected the over two-thirds of the whole sample.
We found that the people who classified a low number of galaxies made less reliable classifications. This should be taken into account in future similar projects, especially those involving citizen science. The structures identified for each galaxy are summarized in \tab{ratings}. The statistics of the incidence of the features are presented in \tab{census}. We then investigated the correlation of the incidences of the structures of interest with the mass of the target galaxy and its environment density. The main result is \tab{corrsig} that gives the Spearman rank coefficient of the correlation along with its $p$-value. We compared our results with older publications in \sect{lit}. We found an extremely strong unexpected anticorrelation of the environment density with the occurrence of the foreground pollution by Galactic cirrus  ($p$-value of $3\times10^{-5}$\%), a positive correlation of the mass of the galaxy with the presence of tidal streams, shells and the irregularity of outer isophotes; the correlation with the occurrence tidal tails is not statistically significant. 
{ We found a statistically significant anticorrelation between the environmental density of the galaxy and the presence of a peripheral star forming disk.}
A qualitative interpretation of the results is provided in \sect{interp}. Briefly, we suggests that the correlation of environment density and cirrus incidence is due to the perpendicular orientation of the Milky Way disk plane with respect to the local structure of the  cosmic web. More massive galaxies contain more tidal features and disturbances because of their stronger gravitational attraction and stronger tidal forces.   In the future, we plan to look for the correlations of the incidence of the morphological structures presented here with the many other parameters available for the \atlas galaxies.  We also develop software that would substitute the visual detection and classification of morphology in the future large surveys.

\section*{Acknowledgements}
We thank Dr. Ivana Ebrov\'a for valuable comments. S.P. acknowledges support from the New Researcher Program (Shinjin grant No. 2019R1C1C1009600) through the National Research Foundation of Korea.

\clearpage

\bibliographystyle{mnras}
\bibliography{citace}


\appendix
\section{Methods}
\label{app:meth}
The MATLAS images were inspected visually. The participants were provided a web interface with a navigation tool to display the image and a questionnaire to fill in. { Initially,  participants received a brief training  and instructed  how to use the web interface, classify the structures of interest  and  fill  the questionnaire.} The participants were not required to inspect all galaxies. They could change their answers later, e.g. when they gained more experience. Each participant filled out the questionnaire independently of the others. A best-guess strategy was adopted: the participants always chose the most probable option (in contrast to rejecting a null hypothesis) and, if the classified feature did not fit to any of the offered categories exactly, they chose the closest option (e.g., when the feature appeared as an unusual shell) or  the most probable option (e.g., because the faintness of the feature made the classification difficult).

\subsection{Participants}
\label{app:methparc}
\begin{figure}
        \resizebox{\hsize}{!}{\includegraphics{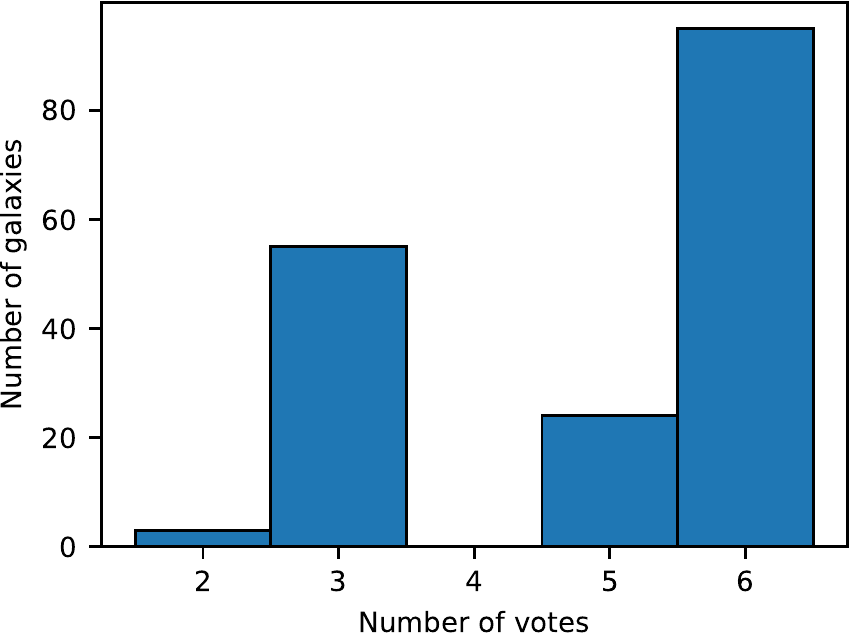}}
        \caption{Histogram of the number of votes per galaxy. The median number of votes per galaxy is six.} 
        \label{fig:votenumbers}
\end{figure}

 There were 15 participants in total. The results we present in this paper are however based just on the votes of six  participants who inspected more than two-thirds of the sample. The other participants inspected usually much less galaxies.  As we show in \app{experience}, the votes of the six experienced participants agree with each other better than if the votes of all participants are considered.  
 A histogram of the number of votes per galaxy that we used is shown in \fig{votenumbers}; the median number of votes per galaxy is six and every galaxy was inspected at least twice.  We used 861 votes.
 
\subsection{Navigation tool}
\label{app:navigator}
\begin{figure*}
	\centering
	\includegraphics[width=17cm]{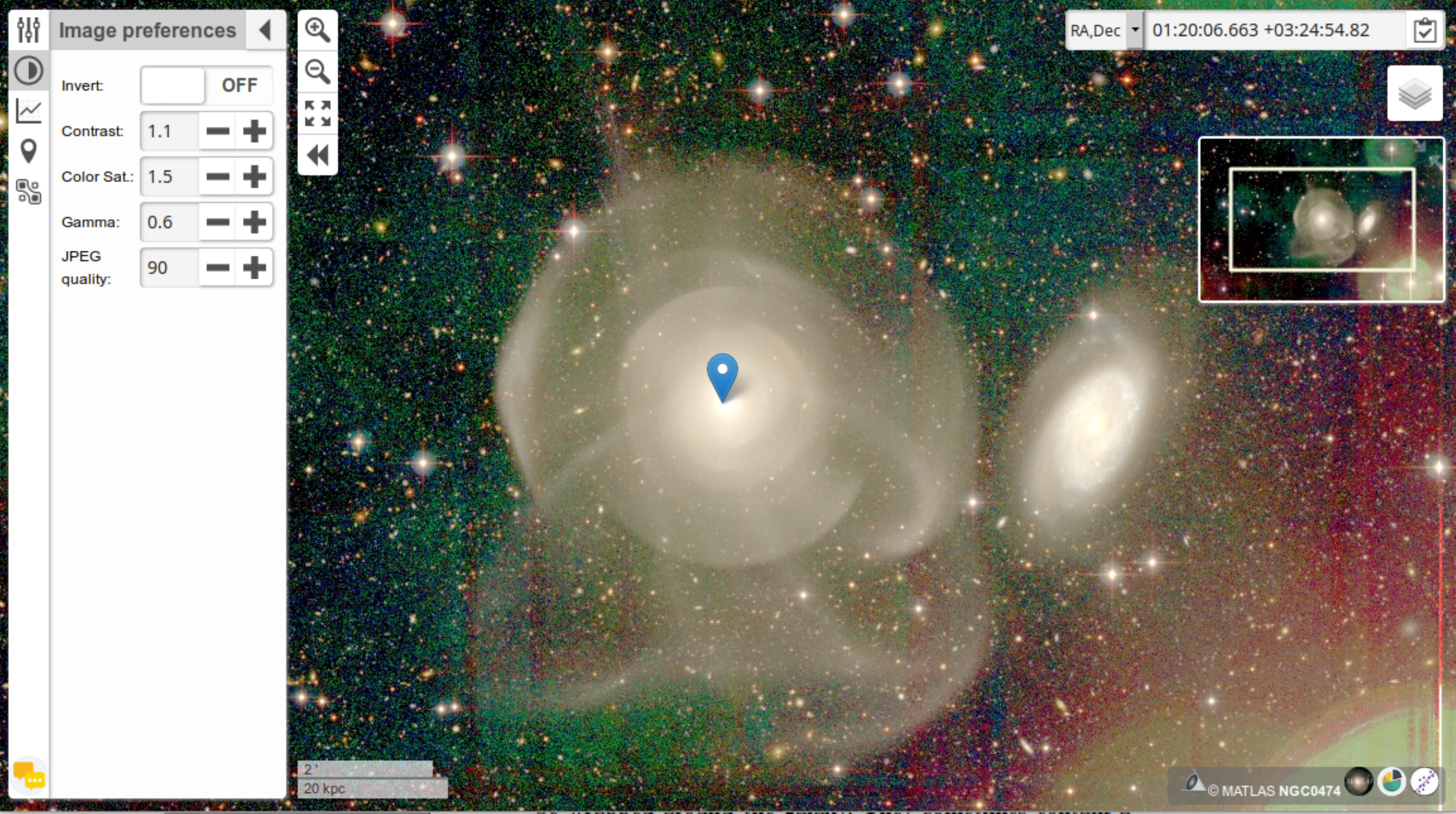}
	\caption{Navigation tool used for the classifications, here displaying the galaxies NGC\,474 (left) and NGC\,470 (right). Accessible at \url{  http://obas-matlas.u-strasbg.fr}. }
	\label{fig:navigator}
\end{figure*}

The navigation tool (\fig{navigator}) used for the classification is based on the VisiOmatic  web client \citep{Bertin15} which is based on the Leaflet Javascript library and the IIPImage server. The tool  allowed us to visualize and navigate and zoom in/out through large science images from remote locations. 
It was customized for the need of the project.

The navigation tool allowed us to inspect the data in many ways.
After opening the web page with the tool (\fig{navigator}),  a composite RGB image was shown.  The RGB image planes were  arranged such that the Red channel contained the $i'$ band data, Green the $r'$ band data, and Blue  the $g'$ band data, when the 3 bands are available.  If images in only two photometric bands were available, namely $g'$ and $r'$, the Green plane was an average of them.  To enhance the low surface brightness features, an arcsinh intensity scaling was applied.  Additional parameters could be set to adjust the nonlinear intensity scaling of the image. The choice of the linear scaling  led for many galaxies to posterization of the image which could have precluded us from detecting some structures. There were also available color index maps that were particularly useful for detecting dust patches and regions of recent star formation. Mono-band surface brightness maps  scaled in mag\,arcsec$^{-2}$ could also be selected.
All images could be zoomed, panned or changed intensity and color saturation levels. We found that the features of interest were visible best if we inspected a grayscale image of only one of the available bands and adjusted the contrast. On the contrary, seeing a faint structure in several photometric bands allowed us to confirm that the object was real. The color images were particular useful when identifying the reflections around bright objects that are the brightest in the $r'$ band filter and appeared as green in the RGB images. 
The images reached the best signal-to-noise ratio in the $g'$ band while the $i'$ band images suffered the least of the parasitic reflections. The visibility of very faint structures depended on the magnification of the image. A tool was available to display the light/color profiles in two directions along a line.  For some galaxies, there were available residual images obtained by subtraction of the smooth model of the galaxy constructed by Galfit \citep{penggalfit}.

Finally, further information on the properties of the galaxies in the \atlas catalog, including massive late-type  galaxies from the parent sample, was provided in displayed labels. These included magnitude, distance and radial velocity. Furthermore, objects in the field  with available SDSS data could be visualized. 

All the on-line images  and the navigation tool that were used in this study  is made publicly available at  \url{  http://obas-matlas.u-strasbg.fr/}.  Examples can be seen in Appendix~\ref{app:img}.

\subsection{Questionnaire}
Here we give a more detailed description of the questionnaire. 

\begin{itemize}
	\item Participants were first asked about the presence of shells, streams or tails around the galaxy. For each type of structure, the available answers were: No, Likely, Yes, and Unsure. The number of features for every category had to be indicated.
	The structures classified here as {\it tails} are defined as elongated features directly attached  to the target galaxy and that appear consistent with having been formed in an ongoing interaction or a recent  merger. The stellar mass of the companion in the interaction has to be similar to or even higher than that of the target galaxy. 	Depending on the morphology of the interacting partners and encounter stage, a tail could appear as an antennae-like structure  similar to the ones observed in the prototypical advanced merger NGC\,7252 (``The { Atoms} for Peace'') -- the merged companion was a spiral galaxy -- or a   plume-like structure with a thickness  comparable to the size of the body of the target galaxy -- the accreted massive companion was an elliptical galaxy or the target ETG is currently interacting with the still existing massive companion.

	The structures classified here as {\it streams} are 
	elongated morphological structures which appear consistent with accreting a companion of a lower mass. They are usually thinner than the tails and can be wrapped around the galaxy. Some of them are detached from the target galaxy. They sometimes contain an embedded candidate for the surviving core of the disrupted galaxy.
	
	{\it Shells} are arc-like sharp-edged features whose centers coincide with the center of the host galaxy. They are usually interpreted as results of radial mergers. The life-times of streams, tails and shells  were estimated in \citet{mancillas19}. We note that dominant formation mechanisms of these structures might be different in the context of modified gravity, where a large fraction of them is expected to arise from non-merging galaxy flybys  \citep{bil18,bil19d}.
	\item The next question was about the shape of the outermost detectable isophote.  The available answers were: Regular, Asymmetric, Disturbed, or Unsure. { By  outermost isophote, we refer to the one which by eye seems unaffected by any artefacts or background fluctuations, and is therefore considered as reliable.  Its level may thus vary from one field to the other and range between about 27 and 29\,mag\,arcsec$^{-2}$ in $g^\prime$.  However, the asymmetries were generally visible even on the brighter isophotes.    The Regular type of the outer isophotes} refers to isophotes that are characterized well by an ellipse possibly with a boxy or disky modulation. The Asymmetric type showed only mild deviations form the regular shape possibly forming an S-shape (a warp), an ovoidal shape, or the ellipse contained a single mild bump. 	Most of the asymmetric isophotes might have been produced by distant or old interactions. On the contrary, the Disturbed isophotes are characterized by a complex shape signifying a recent or strong interaction. We note that if the galaxy contained a very large number of streams or shells, they could not be recognized individually but would instead appear as a blended irregular halo. The most frequent reason for classifying the shape of the outermost isophote as Unsure were either a poor background subtraction caused by the presence of a nearby bright object or that the outermost isophotes overlapped with another large nearby galaxy.
	\item  The presence of dust  was rated on the scale: No, Weak, Strong and Unsure. Dust appears in true color images as darker spots or lanes overlaid over the stellar light of the galaxy. Dust lanes were usually found in the centers of galaxies, but the depth of the images allowed us to identify them even further out. We might have missed the dust lanes at galaxy centers in the case of galaxies that have saturated centers in our deep images. In a few cases, the logged dust lanes were perhaps not in the target galaxy but they were parts of tidal features of a companion galaxy that were overlaid. 
	\item The presence of { a peripheral disk} around the galaxy could be described with one of the following answers: No, Yes or Unsure. { These  disks of stars appear noticeably bluer than the red centers of the galaxies. They can form spiral arms and/or have a clumpy structure.  In some cases, the star forming regions appear to  have an external origin, being captured from a spiral galaxy companion. Because of their faintness, it is unclear whether the peripheral disks extend to the centers of the galaxies. The name ``peripheral'' means that they are best  observed at the outskirts of the galaxies, beyond the main old stellar halo.}
	\item The presence of features usually induced by secular evolution, i.e. internal bars, rings, or spiral arms (with no star formation)   could be rated as No, Weak, Strong and Unsure.  For the galaxies observed edge-on, the presence of  an X-structure was the criterion to estimate the presence of a bar. Saturated regions in the center might have hidden weak bars. We note that bars can also form during galaxy interactions.
	\item The presence of features disturbing the classification had then to be commented, starting with the presence of Galactic cirrus. Their presence and strength  were evaluated as either No, Weak, Strong or Unsure.   Cirri are identified as  filamentary structures which usually come in groups, forming parallel bands or stripes.
	Cirri superimposed on the inspected galaxy could possibly be confused with tidal debris. We annotated also the cirri being further out from the target galaxy but likely affecting the background subtraction decreasing the visibility of the faint structures indirectly.  The large field of view of  MegaCam  allows us to directly identify cirri in the optical images without having to look for infrared images.


\end{itemize}

\subsection{Summarizing the answers, numerical ratings}
\label{app:summarizing}
Since every question was answered several times for every galaxy, once per participant, it was necessary to summarize all the  answers  into one final one. We call the result the rating.  In order to define it, we first assigned numerical values to the answers. For tails, steams, shells, the answers [No; Likely; Yes; Unsure] were respectively converted to [0; 1; 2; -1]; similarly for outer isophotes the answers [Regular; Asymmetric; Disturbed; Unsure] were converted to [0; 1; 2; -1]; {{and  for dust lanes, bars and cirrus [No; Weak; Strong; Unsure]  to [0; 1; 2; -1];  finally for polluting halos or { peripheral disks},  the answers [No; Yes; Unsure] were respectively converted to [0; 1; -1].}} Then we excluded for every galaxy the -1 answers motivated by our best-guess strategy. The rating of the presence of the given feature in the galaxy was defined as the answer that got most votes.  In the ``draw case'' that several answers got the same highest number of votes, the result was the average of their numeric values: for example, if the votes were 2 for ``Yes'', 2 for ``Likely'' and 1 for ``No'', the result was $(2\times2+2\times1)/4 = 1.5$. If all answers were -1, then the final rating was -1. {{When evaluating the polluting halos, the target galaxies whose isophotes overlap with another substantial neighbor were rated as 1 directly without voting, since there is rarely doubt about about such cases. Two of us identified such galaxies together.}} It was also necessary to  decide about the number of streams, tails and shells in every galaxy. This was done in a similar way. We first chose the most frequent answers on presence of the feature from the set [Yes, Likely, No]. Then we worked just with votes of the participants who voted for the most frequent answers: for example, if there were 2 votes for ``Yes'', 2 for ``Likely'' and 1 for ``No'', we worked just with the votes of the participants who voted ``Yes'' or ``Likely''. The resulting number of the features was then calculated as the  weighed average of the numbers given by the selected participants; the weights were either [1; 0.5;  0]  for the presence vote [Yes, Likely, No], respectively. If all answers were ``Unknown'' then the result was -1. 


We also used the ``rounded rating''. Here, we rounded the continuous rating described in the previous paragraph to the nearest integer (e.g., the rating of the presence of streams of 0.3 was rounded to 0). In the case of tidal features, bars, dust lanes and cirrus, the border values of 0.5 and 1.5 were rounded respectively to 0 or 2 in order to minimize the number of objects sorted in the intermediate category. In the case of {{polluting halos}} and { peripheral disks}, a continuous rating of 0.5 was rounded to 1.

\begin{figure*}
	\centering
	\includegraphics[width=17cm]{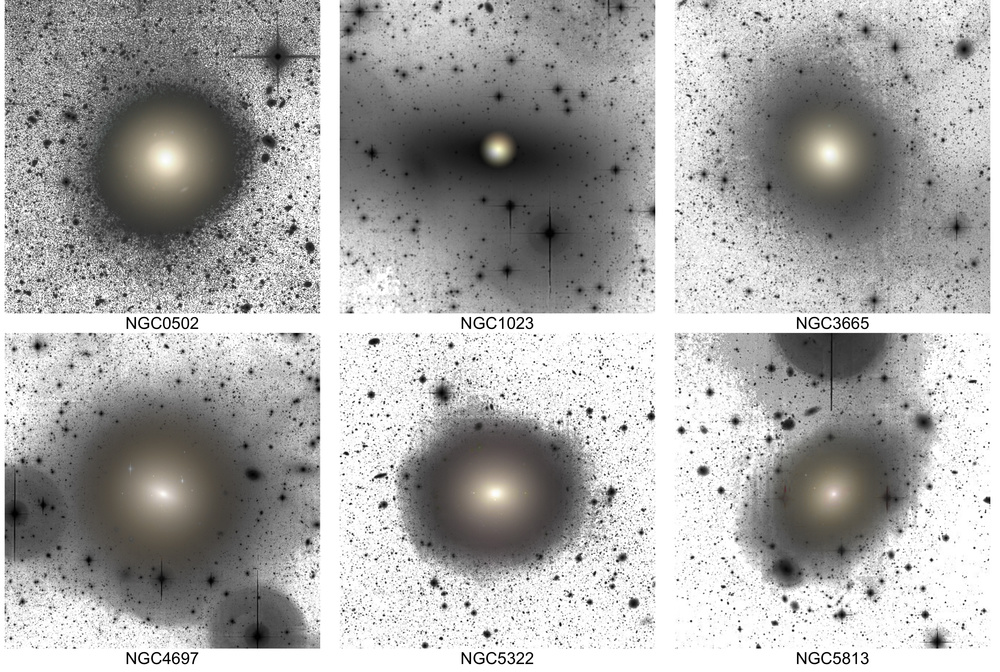}
		\caption{{ Examples of galaxies whose outer isophotes were rated as asymmetric, but that are devoid of prominent tidal features. The asymmetric classification signifies that the outer isophotes deviate from ellipses just mildly, according to the judgement of the participants.} }
	\label{fig:mosasymmetric}
\end{figure*}

\begin{figure*}
	\centering
	\includegraphics[width=17cm]{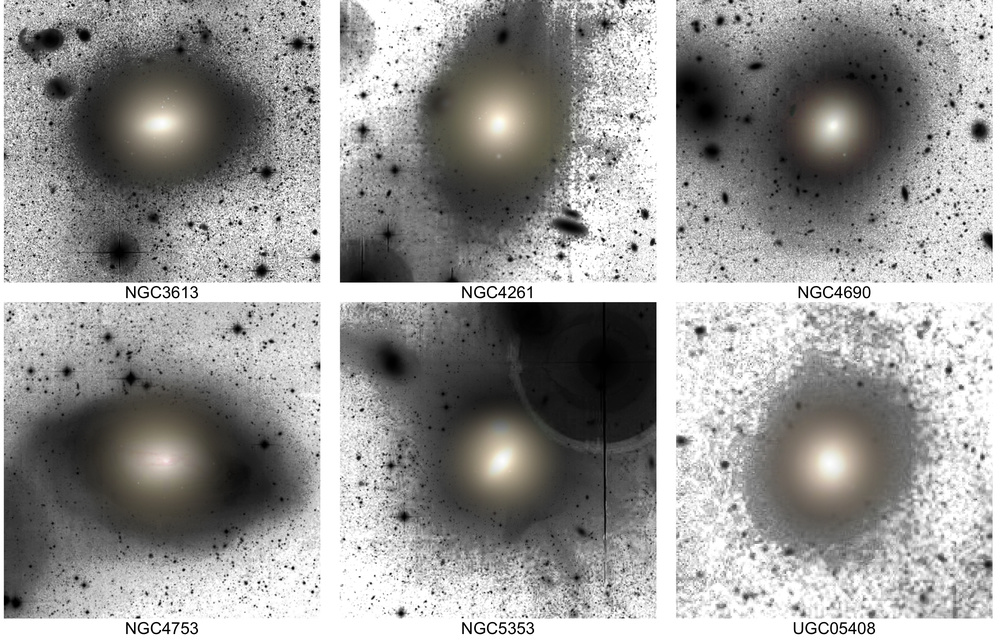}
		\caption{{ Examples of galaxies whose outer isophotes were rated as disturbed,  but that are devoid of prominent tidal features. The disturbed classification signifies that the outer isophotes deviate from ellipses significantly, according to the judgement of the participants.} }
	\label{fig:mosdisturbed}
\end{figure*}

\begin{figure*}
	\centering
	\includegraphics[width=17cm]{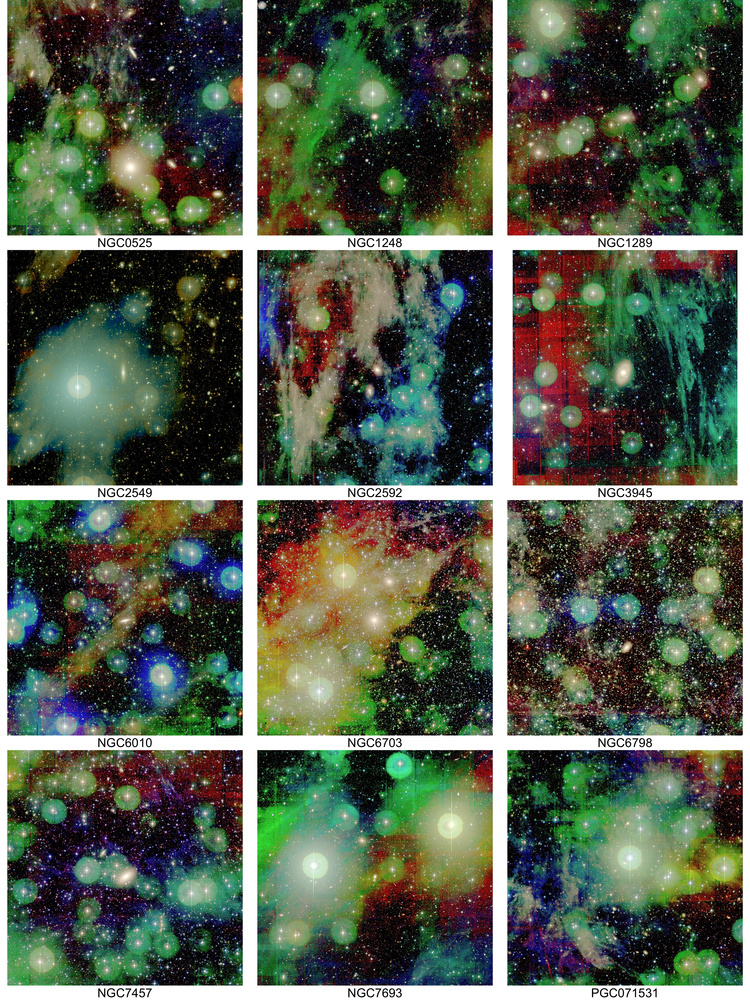}
		\caption{Examples of galaxies where the identification of the faint structure was complicated by the presence of strong Galactic cirri.}
	\label{fig:moscirrus}
\end{figure*}

{{Our classifications could potentially have been affected by the so-called halos. We thus had to identify the galaxies that are affected by halos in order to assess their influence later. The first type of halos are ghost reflections, instrumental artifacts, appearing around bright stars, see, e.g., Figures~\ref{fig:mosasymmetric}--\ref{fig:moscirrus}. They can be recognized easily because all bright stars in the image show a reflection an all reflections have the same size and very similar shapes. The reflections have been described in detail in \citet{Karabal17}.  The second type of polluting halos are other galaxies close to the galaxy in question. Even they can be overlaid over tidal features a prevent their detection. We therefore considered the neighbors, whose isophotes overlap with the isophotes of the targets, as polluting halos. We logged only the halos, of either type, that could have plausibly caused a misclassification of the galaxy of interest. The polluting halos were identified by two people working together. Identifying polluting halos is generally easier and less subjective than, e.g.,  identifying  tidal features. If a galaxy was polluted by a halo, it was assigned a numerical rating of halos of 1, and 0 in the opposite case.}}

\section{The effect of the experience of the participants}
\label{app:experience}
\begin{figure*}
\centering
	\includegraphics[width=17cm]{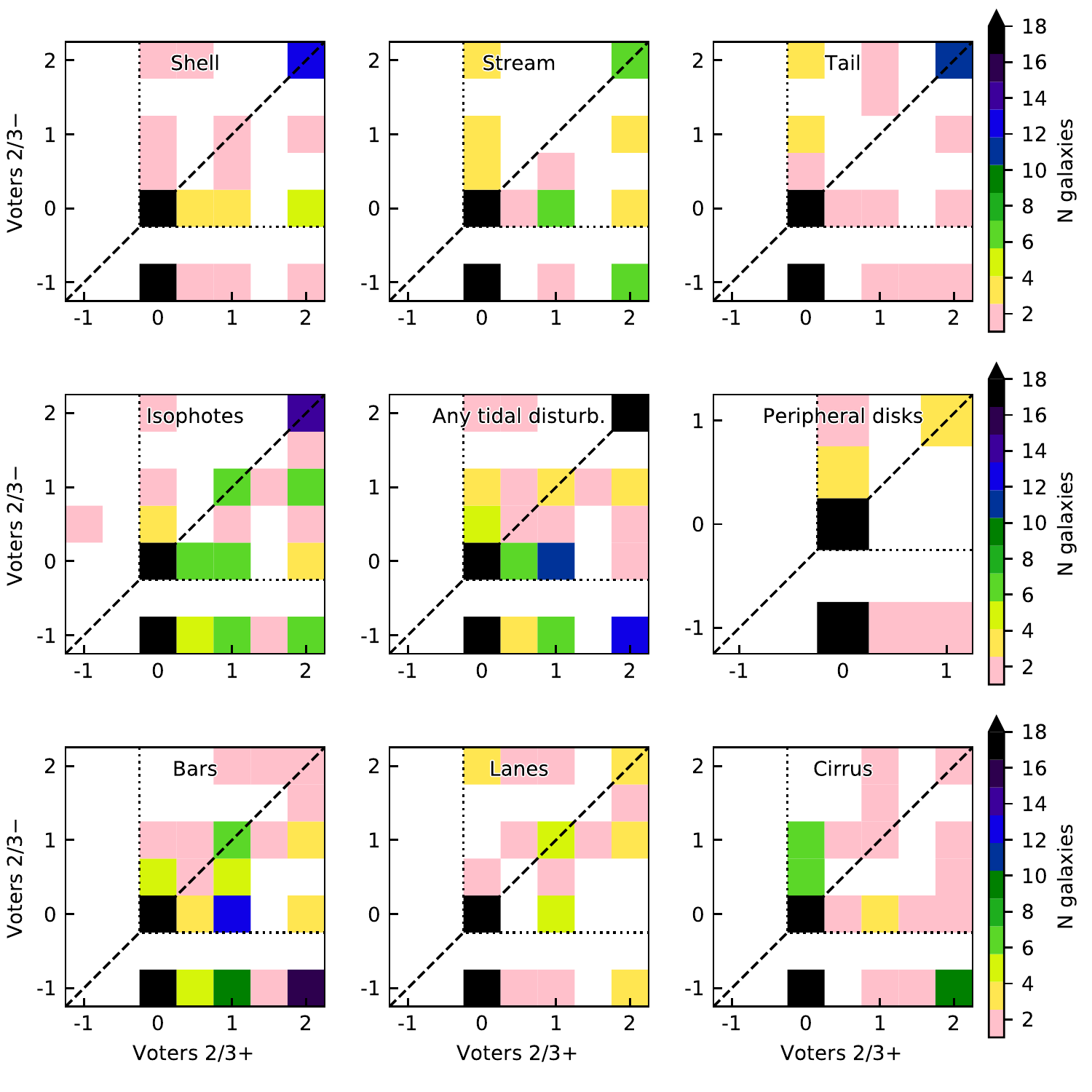}
	\caption{Comparison of the ratings based on the  votes of the participants who inspected over {{two-thirds of}} galaxies (horizontal axis), and the ratings based on the votes of the participants who inspected a lower number of galaxies (vertical axis). The ratings of -1 mostly mean that there are no votes for the galaxy.}
	\label{fig:expvsnonexp}
\end{figure*}

\begin{figure*}
	\centering
	\includegraphics[width=17cm]{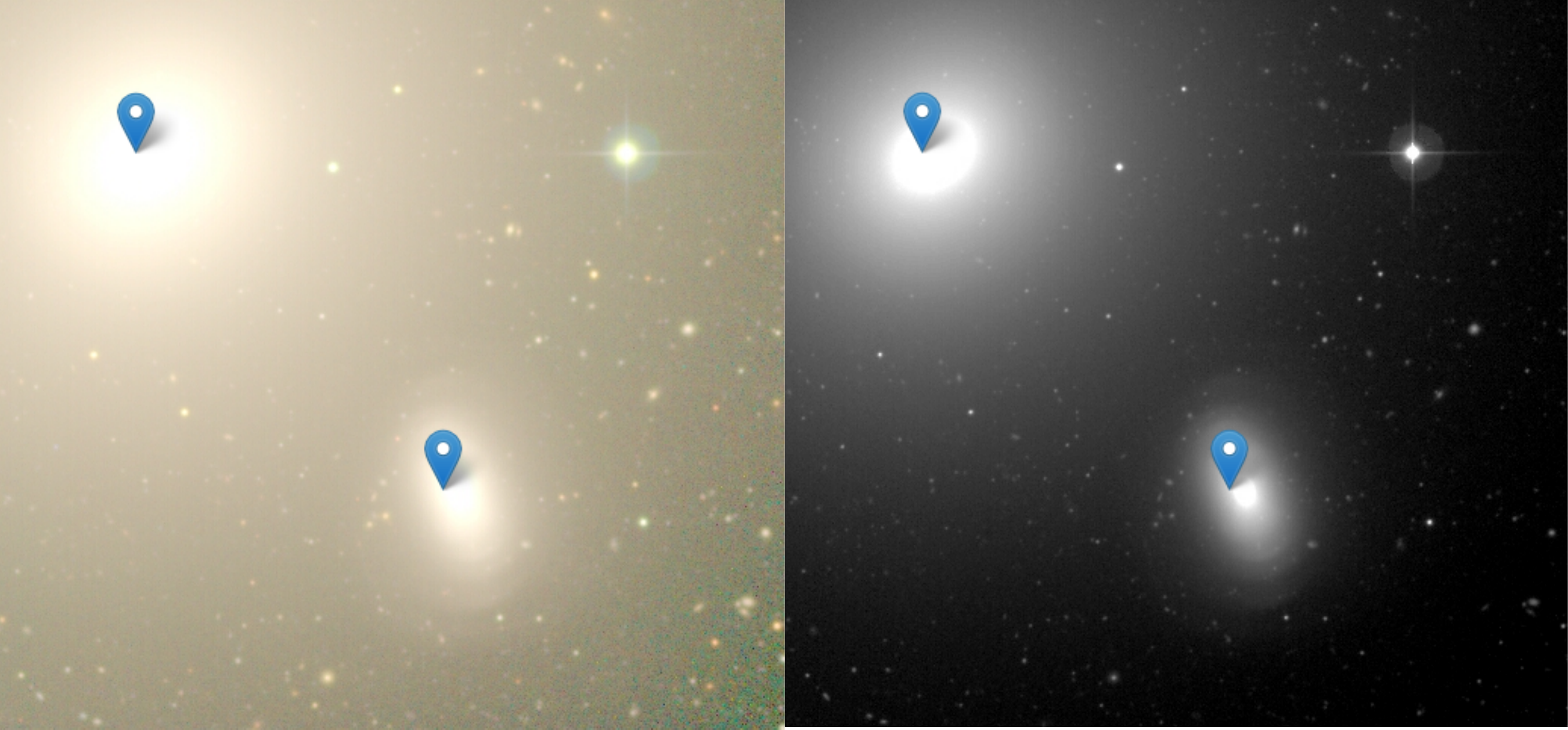}
	\caption{Shells of NGC\,3605 (the smaller galaxy) are an example of a feature that was difficult to detect for the less experienced participants. Left panel shows the default view that was displayed to everyone. The right panel demonstrates that the shells are visible clearly after increasing the contrast of the image in the navigation tool. }
	\label{fig:misclass}
\end{figure*}

In  Figure~\ref{fig:expvsnonexp} we compared the adopted rating to the  rating based only on the votes of the less experienced participants {{, i.e. those who inspected less than two thirds of the sample}}.  We examined the cases where the two ratings differed completely, i.e. where one came out to have the maximum value while the other came out 0. In most cases the difference in rating can be accounted to classifying a well-visible feature in another category. The less experienced classifiers also had difficulties with detecting low-contrast features in a few cases.  Figure~\ref{fig:misclass} presents an example of a galaxy where shells were detected according to the standard rating but were not according to the rating by the less experienced group. The shells were likely missed because the galaxy with shells, NGC\,3605, is seen overlaid over the body of a bright and large neighbor, NGC\,3607, and therefore the contrast of the shells is reduced. The right panel of the figure shows that the contrast can be set in the navigator tool such that the shells become clearly visible. Those participants who have had inspected just a few galaxies before this one might have also confused the two neighboring galaxies.

We noted in Figure~\ref{fig:expvsnonexp} the following:  1) The more experienced classifiers were more positive about the presence of shells, streams, disturbed isophotes, bars and cirrus. This is likely because the less experienced contributors missed the low-contrast features.  
   2) The less experienced contributors were more positive about the presence of tidal tails probably because of the confusion with streams. 
   3) The two groups of participants agree well about the presence of dust lanes, a feature easily detectable. 
   4) The less experienced voters were more positive about the presence of { peripheral disks}. The confusion happened in this case usually because of disturbing reflections of nearby bright stars or because of a change of color of the galaxy due to the PSF effects.
   5) The two groups agree well when there is no feature or when the feature is prominent.

We further explored  whether excluding the less experienced participants leads to more consistent results. For each galaxy and each feature class we calculated the standard deviation of the votes (after the conversion to a numerical value) on the presence of the feature within the group of the more experienced voters and within the group of all participants together. Then we calculated the result for each feature type as a weighed average of the scatters for individual galaxies, while the weights were the number of votes for the given galaxy. The comparison of the average scatters for the given feature for the group of experienced participants and the group of all contributors is presented in \tab{scatter}. The scatter between the two groups is mostly lower for the experienced participants or it is the same. We note that 59 galaxies, i.e. 33\% of the sample, had votes only by the more experienced voters. Finally, we present in \tab{commoncensus} the census based on the votes of all participants. One can note by comparison with the main census  \tab{census} that the two are consistent.

\begin{table}
\caption{Comparison of the scatter in the in the default rating, based on the  votes of the participants who inspected over {{two-thirds of all}} galaxies, and the rating based on the votes of all participants.}
\label{tab:scatter}
\begin{tabular}{lp{2.3cm}p{2.5cm}}
\hline\hline
Structure & More experienced voters only & All voters           \\
\hline
Shells & 0.23 & 0.25 \\ 
Streams & 0.32 & 0.35 \\ 
Tails & 0.18 & 0.23 \\ 
Outer isophotes & 0.43 & 0.43 \\ 
Peripheral disks & 0.087 & 0.092 \\ 
Secular features & 0.45 & 0.45 \\ 
Dust & 0.25 & 0.27 \\ 
Cirrus & 0.37 & 0.38 \\ 
Halos & 0.27 & 0.27 \\  
\hline
\end{tabular}
\end{table}

\begin{table*}
	\caption{Census of the classified structures in the whole sample according to the rounded rating based on the votes of all participants (i.e., including also the less experienced ones) in percent.}
	\label{tab:commoncensus}
	\centering
	\begin{tabular}{lllll}
		\hline\hline
Shells  & no: $88 \pm 7$   & likely: $3 \pm 1$   & yes: $10 \pm 2$   & unknown: $0 \pm 0$ \\ 
Streams  & no: $87 \pm 7$   & likely: $3 \pm 1$   & yes: $10 \pm 2$   & unknown: $0 \pm 0$ \\ 
Tails  & no: $87 \pm 7$   & likely: $2 \pm 1$   & yes: $11 \pm 3$   & unknown: $0 \pm 0$ \\ 
Outer isophotes  & regul.: $71 \pm 6$   & asym.: $11 \pm 2$   & disturb.: $18 \pm 3$   & unsure: $0 \pm 0$ \\ 
Tails or streams  & no: $80 \pm 7$   & likely: $2 \pm 1$   & yes: $19 \pm 3$   & unknown: $0 \pm 0$ \\ 
Shells or streams or tails  & no: $76 \pm 7$   & likely: $3 \pm 1$   & yes: $21 \pm 3$   & unknown: $0 \pm 0$ \\ 
Any tidal disturbance  & no: $66 \pm 6$   & likely: $8 \pm 2$   & yes: $26 \pm 4$   & unknown: $0 \pm 0$ \\ 
Bars  & no: $73 \pm 6$   & weak: $12 \pm 3$   & strong: $15 \pm 3$   & unsure: $0 \pm 0$ \\ 
Dust lanes  & no: $86 \pm 7$   & weak: $6 \pm 2$   & strong: $8 \pm 2$   & unsure: $0 \pm 0$ \\ 
Peripheral disks  & no: $96 \pm 7$ &    & yes: $4 \pm 1$   & unsure: $0 \pm 0$ \\ 
Cirrus  & no: $85 \pm 7$   & weak: $4 \pm 1$   & strong: $11 \pm 2$   & unsure: $0 \pm 0$ \\ 
		\hline
	\end{tabular}
\end{table*}


\section{Examples of the classified structures}
\label{app:img}

\begin{figure*}
	\centering
	\includegraphics[width=17cm]{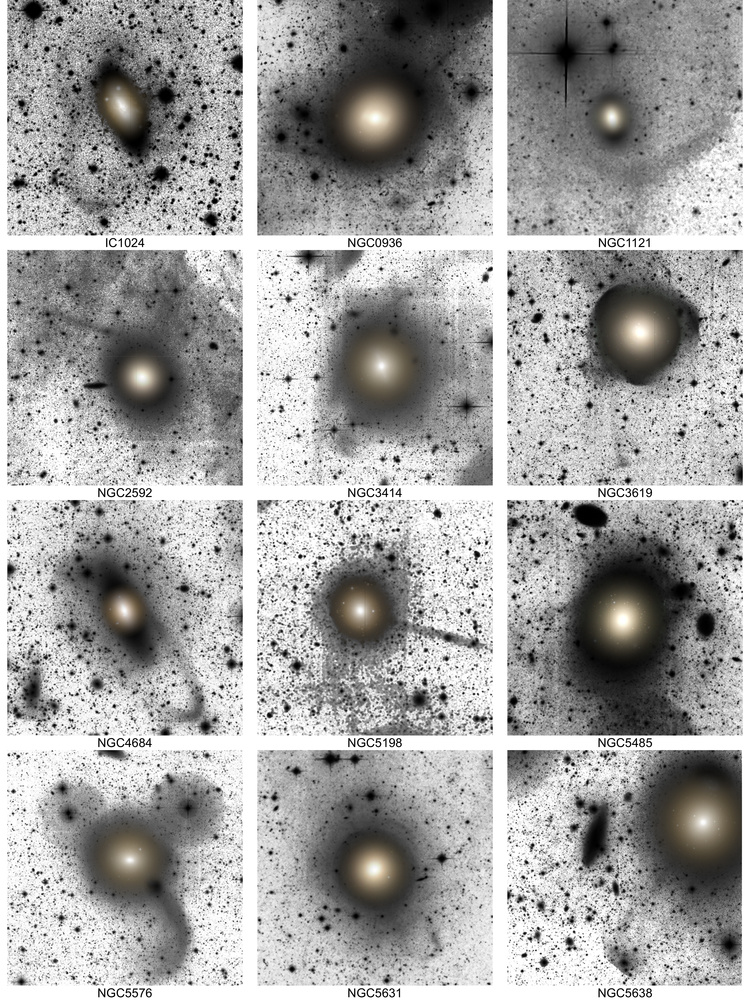}
		\caption{Examples of galaxies that were classified as having streams.}
	\label{fig:mosst}
\end{figure*}

\begin{figure*}
	\centering
	\includegraphics[width=17cm]{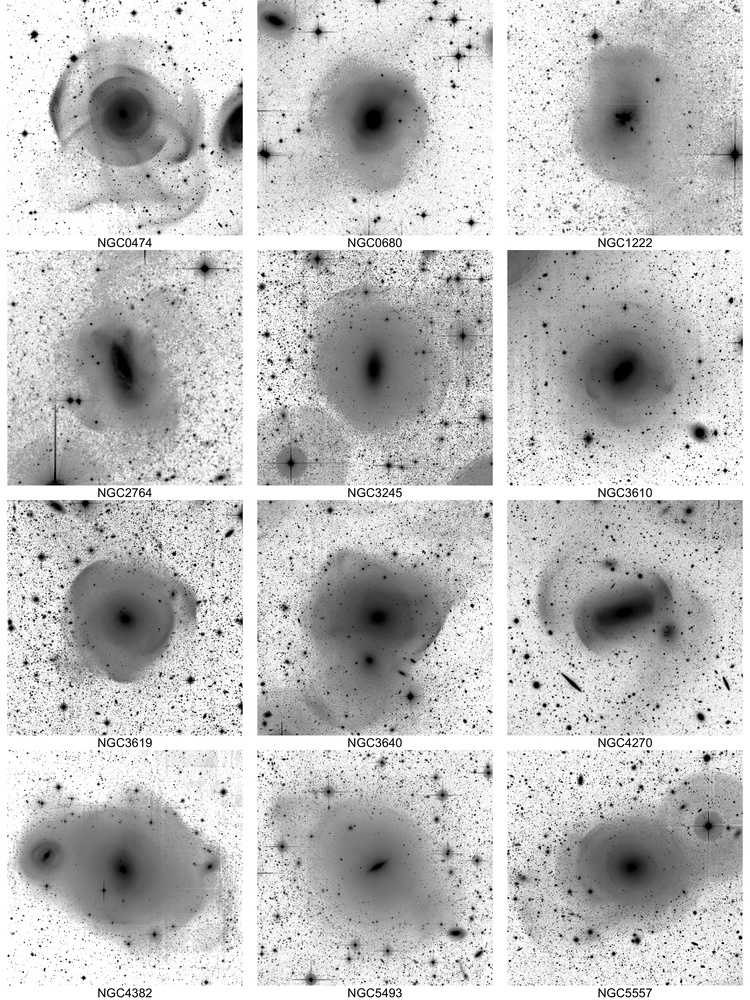}
		\caption{Examples of galaxies that were classified as having shells.}
	\label{fig:mossh}
\end{figure*}

\begin{figure*}
	\centering
	\includegraphics[width=17cm]{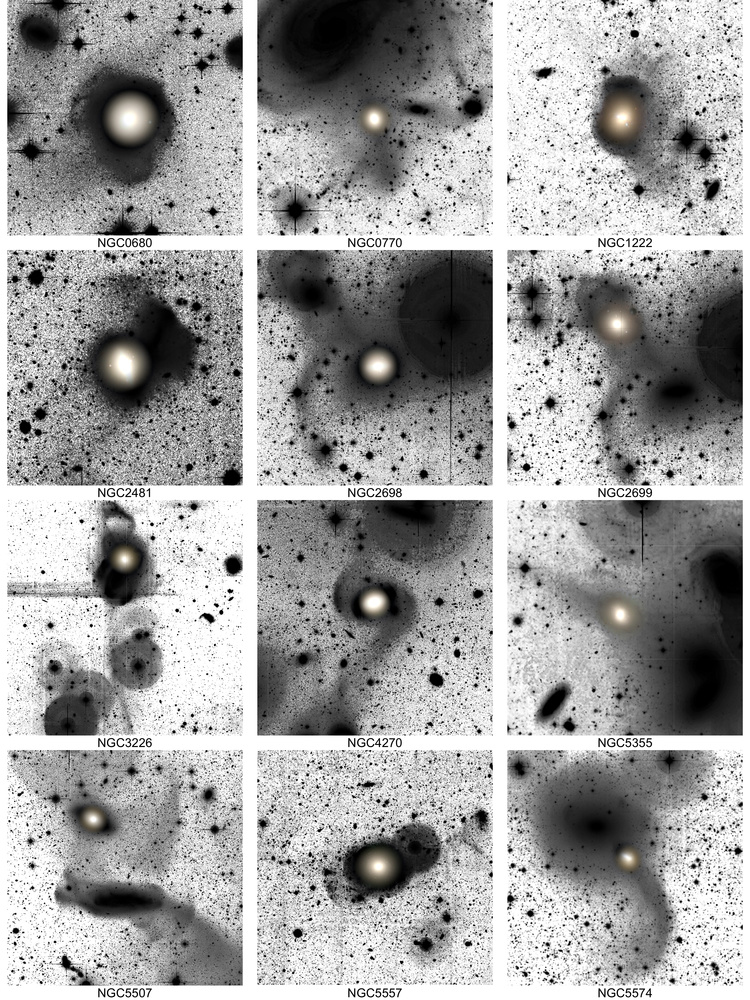}
		\caption{Examples of galaxies that
		were classified as having tidal tails.}
	\label{fig:mostail}
\end{figure*}

\begin{figure*}
	\centering
	\includegraphics[width=17cm]{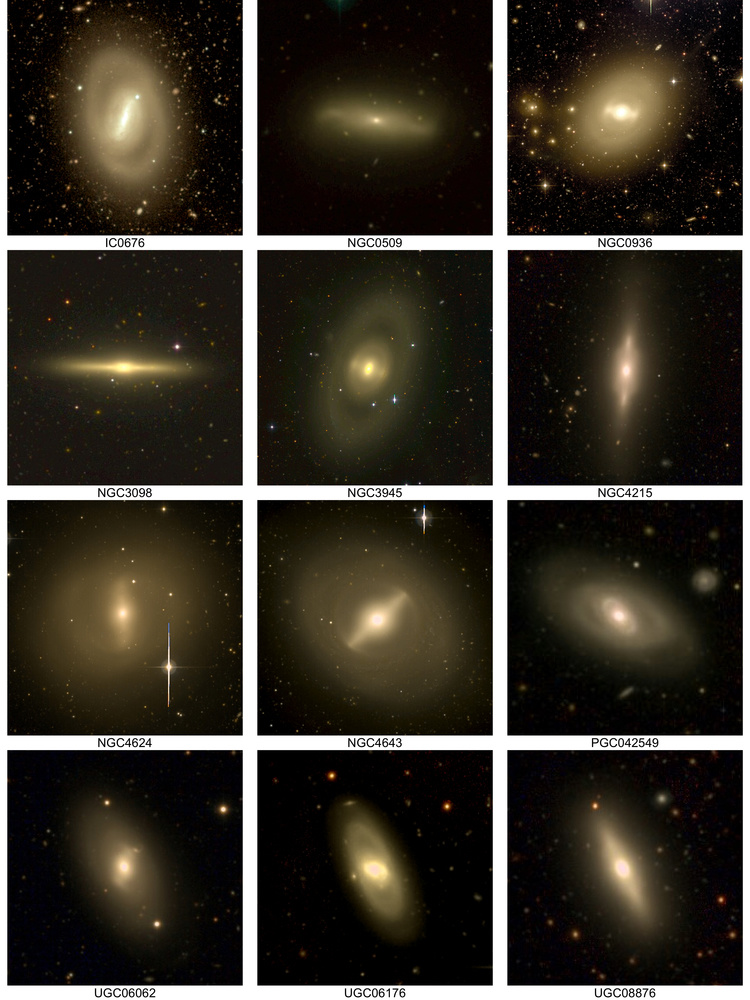}
		\caption{Examples of galaxies that were classified as having bars.}
	\label{fig:mosbars}
\end{figure*}

\begin{figure*}
	\centering
	\includegraphics[width=17cm]{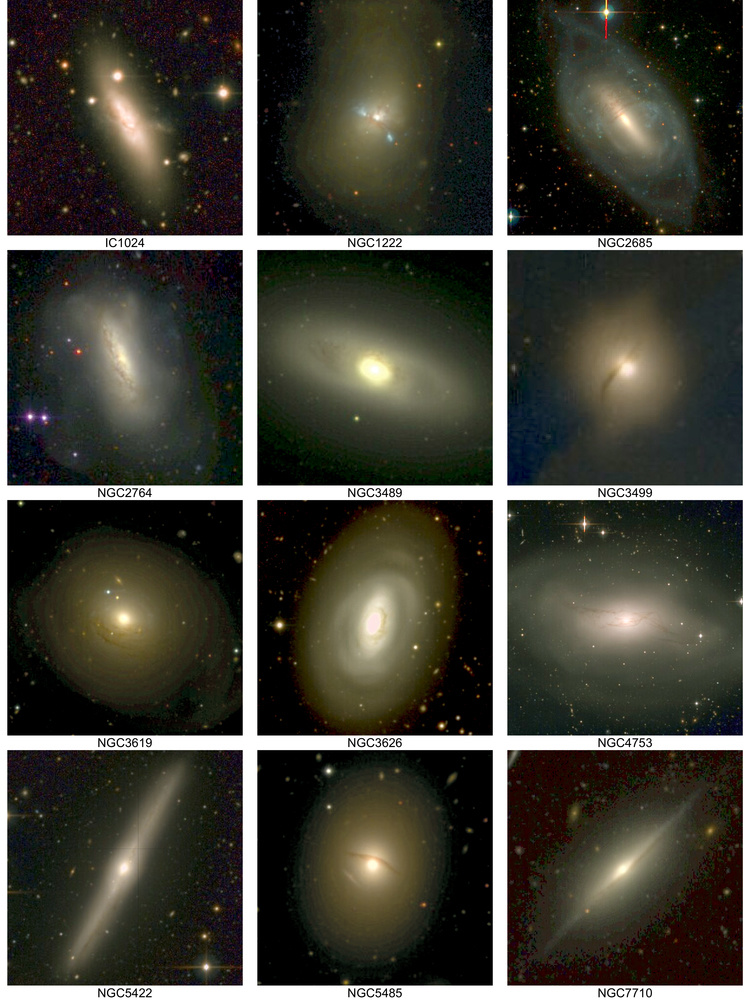}
		\caption{Examples of galaxies that were classified as having dust lanes.}
	\label{fig:moslanes}
\end{figure*}

\begin{figure*}
	\centering
	\includegraphics[width=17cm]{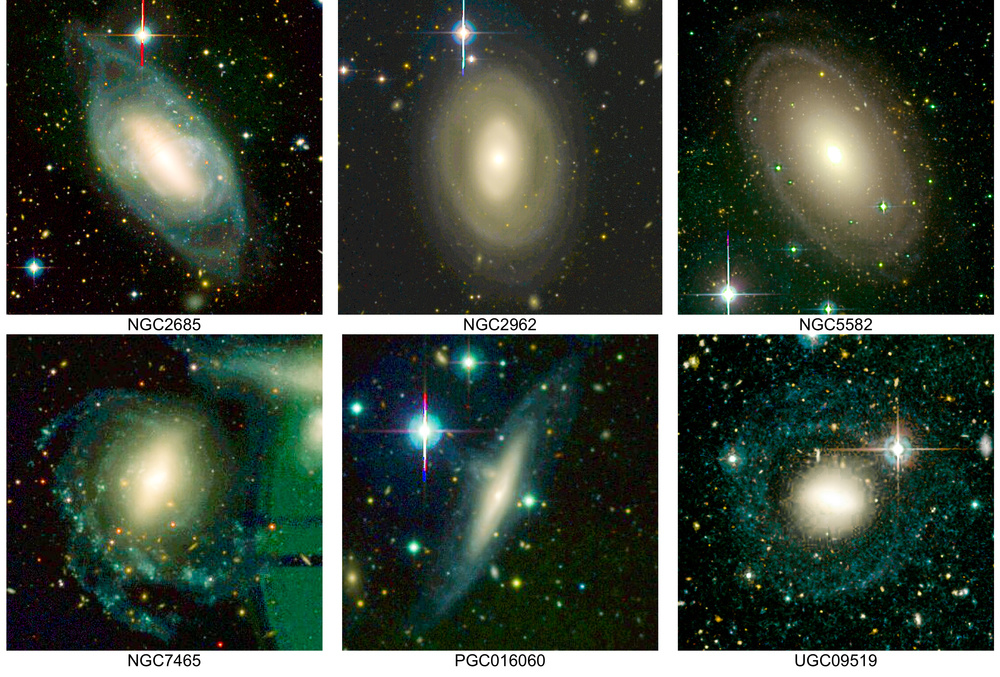}
		\caption{Examples of galaxies that were classified as having { peripheral disks}.}
	\label{fig:mosextsf}
\end{figure*}

\clearpage
\onecolumn
\section{Presence of morphological features in individual galaxies}
\label{app:ratings}
\begin{longtable}{lllllllllll}
\caption{\label{tab:ratings} Ratings and classification codes for all galaxies.}\\
\hline\hline
Name &  Classification code  &  Streams & Shells & Tails & Ext. SF & Out. isoph. & Dust & Bars & Halos & Cirrus \\
\hline 
\endfirsthead
\caption{continued.}\\
\hline\hline
Name &  Classification code  &  Streams & Shells & Tails & Ext. SF & Out. isoph. & Dust & Bars & Halos & Cirrus \\
\hline 
\endhead
\hline
\endfoot
                
               IC0560         & +wb                       & 0.0      & 0.0      & 0.0      & 0.0      & 0.0      & 0.0      & 1.0      & 0.0      & 0.0     \\
IC0598         &                           & 0.0      & 0.0      & 0.0      & 0.0      & 0.0      & 0.0      & 0.0      & 0.0      & 0.0     \\
IC0676         & +pb                       & 0.0      & 0.0      & 0.0      & 0.0      & 0.0      & 0.0      & 2.0      & 0.0      & 0.0     \\
IC1024         & +s+ph+pl                  & 2.0      & 0.0      & 0.0      & 0.0      & 2.0      & 2.0      & 0.0      & 0.0      & 0.0     \\
NGC0448        &                           & 0.0      & 0.0      & 0.0      & 0.0      & 0.0      & 0.0      & 0.0      & 0.0      & 0.0     \\
NGC0474        & +s+r+t+ph-h               & 2.0      & 2.0      & 2.0      & 0.0      & 2.0      & 0.0      & 0.0      & 1.0      & 0.0     \\
NGC0502        & +r+ah                     & 0.0      & 2.0      & 0.0      & 0.0      & 1.0      & 0.0      & 0.0      & 0.0      & 0.0     \\
NGC0509        & +pb                       & 0.0      & 0.0      & 0.0      & 0.0      & 0.0      & 0.0      & 2.0      & 0.0      & 0.0     \\
NGC0516        & +wb                       & 0.0      & 0.0      & 0.0      & 0.0      & 0.0      & 0.0      & 1.0      & 0.0      & 0.0     \\
NGC0524        & +h?+wl-pc                 & 0.0      & 0.5      & 0.0      & 0.0      & -1       & 1.0      & 0.0      & 0.0      & 2.0     \\
NGC0525        & +wb-pc                    & 0.0      & 0.0      & 0.0      & 0.0      & 0.0      & 0.0      & 1.0      & 0.0      & 2.0     \\
NGC0661        &                           & 0.0      & 0.0      & 0.0      & 0.0      & 0.0      & 0.0      & 0.0      & 0.0      & 0.0     \\
NGC0680        & +s+r+t+ph+wl-h-wc         & 2.0      & 2.0      & 2.0      & 0.0      & 2.0      & 1.0      & 0.0      & 1.0      & 1.0     \\
NGC0770        & +t+ah-h-pc                & 0.0      & 0.0      & 2.0      & 0.0      & 1.0      & 0.0      & 0.0      & 1.0      & 2.0     \\
NGC0821        & -pc                       & 0.0      & 0.0      & 0.0      & 0.0      & 0.5      & 0.0      & 0.0      & 0.0      & 2.0     \\
NGC0936        & +s+pb                     & 2.0      & 0.0      & 0.0      & 0.0      & 0.0      & 0.0      & 2.0      & 0.0      & 0.0     \\
NGC1023        & +ah                       & 0.0      & 0.0      & 0.0      & 0.0      & 1.0      & 0.0      & 0.5      & 0.0      & 0.0     \\
NGC1121        & +s                        & 2.0      & 0.0      & 0.0      & 0.0      & 0.0      & 0.0      & 0.0      & 0.0      & 0.0     \\
NGC1222        & +r+t+ph+pl                & 0.0      & 2.0      & 2.0      & 0.0      & 2.0      & 2.0      & 0.0      & 0.0      & 0.0     \\
NGC1248        & -h-pc                     & 0.0      & 0.0      & 0.0      & 0.0      & 0.0      & 0.0      & 0.5      & 1.0      & 2.0     \\
NGC1266        & +wl-wc                    & 0.0      & 0.0      & 0.0      & 0.0      & 0.0      & 1.0      & 0.0      & 0.0      & 1.0     \\
NGC1289        & -pc                       & 0.0      & 0.0      & 0.0      & 0.0      & 0.0      & 0.0      & 0.0      & 0.0      & 1.5     \\
NGC1665        & +wb-wc                    & 0.0      & 0.0      & 0.0      & 0.0      & 0.0      & 0.0      & 1.0      & 0.0      & 1.0     \\
NGC2481        & +t+ph-h-pc                & 0.0      & 0.0      & 2.0      & 0.0      & 2.0      & 0.0      & 0.5      & 1.0      & 1.5     \\
NGC2549        & +wb-pc                    & 0.0      & 0.0      & 0.0      & 0.0      & 0.0      & 0.0      & 1.0      & 0.0      & 2.0     \\
NGC2577        & -wc                       & 0.0      & 0.0      & 0.0      & 0.0      & 0.0      & 0.0      & 0.0      & 0.0      & 1.0     \\
NGC2592        & +s-pc                     & 2.0      & 0.0      & 0.0      & 0.0      & 0.0      & 0.0      & 0.0      & 0.0      & 2.0     \\
NGC2594        & +ah+wb-h-wc               & 0.0      & 0.0      & 0.0      & 0.0      & 1.0      & 0.0      & 1.0      & 1.0      & 1.0     \\
NGC2679        & +wb                       & 0.0      & 0.0      & 0.0      & 0.0      & 0.0      & 0.0      & 1.0      & 0.0      & 0.0     \\
NGC2685        & +t?+d+ah+pl               & 0.0      & 0.0      & 1.0      & 1.0      & 1.0      & 2.0      & 0.5      & 0.0      & 0.0     \\
NGC2695        &                           & 0.0      & 0.0      & 0.0      & 0.0      & 0.0      & 0.0      & 0.0      & 0.0      & 0.0     \\
NGC2698        & +t+ah-h                   & 0.0      & 0.0      & 2.0      & 0.0      & 1.0      & 0.0      & 0.0      & 1.0      & 0.0     \\
NGC2699        & +r+t+ph-h                 & 0.0      & 2.0      & 2.0      & 0.0      & 2.0      & 0.0      & 0.0      & 1.0      & 0.5     \\
NGC2764        & +r+ph+pl                  & 0.0      & 2.0      & 0.0      & 0.0      & 2.0      & 2.0      & 0.0      & 0.0      & 0.0     \\
NGC2768        & +s?+ah+wb                 & 1.0      & 0.0      & 0.0      & 0.0      & 1.0      & 0.0      & 1.0      & 0.0      & 0.0     \\
NGC2778        & -h                        & 0.0      & 0.0      & 0.0      & 0.0      & 0.0      & 0.0      & 0.0      & 1.0      & 0.0     \\
NGC2824        & +d                        & 0.0      & 0.0      & 0.0      & 0.5      & 0.0      & 0.0      & 0.0      & 0.0      & 0.0     \\
NGC2852        & -h-wc                     & 0.0      & 0.0      & 0.0      & 0.0      & 0.0      & 0.0      & 0.0      & 1.0      & 1.0     \\
NGC2859        & +pb                       & 0.0      & 0.0      & 0.0      & 0.0      & 0.0      & 0.0      & 1.5      & 0.0      & 0.0     \\
NGC2880        &                           & 0.0      & 0.0      & 0.0      & 0.0      & 0.5      & 0.0      & 0.0      & 0.0      & 0.0     \\
NGC2950        & +wb                       & 0.0      & 0.0      & 0.0      & 0.0      & 0.0      & 0.0      & 1.0      & 0.0      & 0.0     \\
NGC2962        & +d+wb                     & 0.0      & 0.0      & 0.0      & 1.0      & 0.0      & 0.0      & 1.0      & 0.0      & 0.0     \\
NGC2974        &                           & 0.0      & 0.0      & 0.0      & 0.0      & 0.5      & 0.0      & 0.0      & 0.0      & 0.0     \\
NGC3032        &                           & 0.0      & 0.0      & 0.0      & 0.0      & 0.0      & 0.0      & 0.0      & 0.0      & 0.0     \\
NGC3073        & +ph                       & 0.0      & 0.0      & 0.0      & 0.0      & 1.5      & 0.0      & 0.0      & 0.0      & 0.0     \\
NGC3098        & +pb                       & 0.0      & 0.0      & 0.0      & 0.0      & 0.0      & 0.0      & 2.0      & 0.0      & 0.0     \\
NGC3156        & +wl                       & 0.0      & 0.0      & 0.0      & 0.0      & 0.0      & 1.0      & 0.0      & 0.0      & 0.0     \\
NGC3182        &                           & 0.0      & 0.0      & 0.0      & 0.0      & 0.0      & 0.0      & 0.0      & 0.0      & 0.0     \\
NGC3193        & -h                        & 0.0      & 0.0      & 0.0      & 0.0      & 0.0      & 0.0      & 0.0      & 1.0      & 0.0     \\
NGC3226        & +s?+t+ph+wl-h             & 1.0      & 0.5      & 2.0      & 0.0      & 2.0      & 1.0      & 0.0      & 1.0      & 0.0     \\
NGC3230        & +pb-wc                    & 0.0      & 0.0      & 0.0      & 0.0      & 0.0      & 0.0      & 2.0      & 0.0      & 1.0     \\
NGC3245        & +r-h                      & 0.0      & 2.0      & 0.0      & 0.0      & 0.0      & 0.0      & 0.0      & 1.0      & 0.0     \\
NGC3248        & +wb                       & 0.0      & 0.0      & 0.0      & 0.0      & 0.0      & 0.0      & 1.0      & 0.0      & 0.0     \\
NGC3301        &                           & 0.0      & 0.0      & 0.0      & 0.0      & 0.0      & 0.0      & 0.5      & 0.0      & 0.0     \\
NGC3377        & -h                        & 0.0      & 0.0      & 0.0      & 0.0      & 0.0      & 0.0      & 0.0      & 1.0      & 0.0     \\
NGC3379        & +r?-h                     & 0.0      & 1.0      & 0.0      & 0.0      & 0.0      & 0.0      & 0.0      & 1.0      & 0.0     \\
NGC3384        & +wb-h                     & 0.0      & 0.0      & 0.0      & 0.0      & 0.5      & 0.0      & 1.0      & 1.0      & 0.0     \\
NGC3400        & +pb                       & 0.0      & 0.0      & 0.0      & 0.0      & 0.0      & 0.0      & 2.0      & 0.0      & 0.0     \\
NGC3412        &                           & 0.0      & 0.0      & 0.0      & 0.0      & 0.0      & 0.0      & 0.5      & 0.0      & 0.0     \\
NGC3414        & +s+t?+ph+wb               & 2.0      & 0.5      & 1.0      & 0.0      & 2.0      & 0.0      & 1.0      & 0.0      & 0.0     \\
NGC3457        & +wl                       & 0.0      & 0.0      & 0.0      & 0.0      & 0.0      & 1.0      & 0.0      & 0.0      & 0.0     \\
NGC3458        &                           & 0.0      & 0.0      & 0.0      & 0.0      & 0.0      & 0.0      & 0.5      & 0.0      & 0.0     \\
NGC3489        & +pl+wb                    & 0.0      & 0.0      & 0.0      & 0.0      & 0.0      & 1.5      & 1.0      & 0.0      & 0.0     \\
NGC3499        & +ah+pl                    & 0.0      & 0.0      & 0.0      & 0.0      & 1.0      & 2.0      & 0.0      & 0.0      & 0.0     \\
NGC3522        &                           & 0.0      & 0.0      & 0.0      & 0.0      & 0.0      & 0.0      & 0.0      & 0.0      & 0.0     \\
NGC3530        &                           & 0.0      & 0.0      & 0.0      & 0.0      & 0.0      & 0.0      & 0.0      & 0.0      & 0.0     \\
NGC3599        & +wb                       & 0.0      & 0.0      & 0.0      & 0.0      & 0.0      & 0.0      & 1.0      & 0.0      & 0.0     \\
NGC3605        & +r+ph-h                   & 0.0      & 2.0      & 0.0      & 0.0      & 2.0      & 0.0      & 0.0      & 1.0      & 0.0     \\
NGC3607        & +ph+wl-h                  & 0.0      & 0.0      & 0.5      & 0.0      & 2.0      & 1.0      & 0.0      & 1.0      & 0.0     \\
NGC3608        & +r?+ph-h                  & 0.0      & 1.0      & 0.0      & 0.0      & 1.5      & 0.0      & 0.0      & 1.0      & 0.0     \\
NGC3610        & +r+ph+wb                  & 0.0      & 2.0      & 0.0      & 0.0      & 2.0      & 0.0      & 1.0      & 0.0      & 0.0     \\
NGC3613        & +s+r?+ph                  & 2.0      & 1.0      & 0.0      & 0.0      & 2.0      & 0.0      & 0.0      & 0.0      & 0.0     \\
NGC3619        & +s+r+t?+ph+pl             & 2.0      & 2.0      & 1.0      & 0.0      & 2.0      & 2.0      & 0.0      & 0.0      & 0.0     \\
NGC3626        & +pl+wb                    & 0.0      & 0.0      & 0.0      & 0.0      & 0.0      & 2.0      & 1.0      & 0.0      & 0.0     \\
NGC3630        & +wb                       & 0.0      & 0.0      & 0.0      & 0.0      & 0.0      & 0.0      & 1.0      & 0.0      & 0.0     \\
NGC3640        & +s+r+ph-h-wc              & 2.0      & 2.0      & 0.0      & 0.0      & 2.0      & 0.0      & 0.0      & 1.0      & 1.0     \\
NGC3641        & +ph-h-wc                  & 0.0      & 0.0      & 0.0      & 0.0      & 2.0      & 0.0      & 0.0      & 1.0      & 1.0     \\
NGC3648        &                           & 0.0      & 0.0      & 0.0      & 0.0      & 0.0      & 0.0      & 0.0      & 0.0      & 0.0     \\
NGC3658        &                           & 0.0      & 0.0      & 0.0      & 0.0      & 0.0      & 0.0      & 0.0      & 0.0      & 0.0     \\
NGC3665        & +ah                       & 0.0      & 0.0      & 0.0      & 0.0      & 1.0      & 0.5      & 0.0      & 0.0      & 0.5     \\
NGC3674        &                           & 0.0      & 0.0      & 0.0      & 0.0      & 0.0      & 0.0      & 0.0      & 0.0      & 0.0     \\
NGC3694        &                           & 0.0      & 0.0      & 0.0      & 0.0      & 0.0      & 0.0      & 0.0      & 0.0      & 0.0     \\
NGC3757        & +wb                       & 0.0      & 0.0      & 0.0      & 0.0      & 0.0      & 0.0      & 1.0      & 0.0      & 0.0     \\
NGC3796        &                           & 0.0      & 0.0      & 0.0      & 0.0      & 0.0      & 0.0      & 0.5      & 0.0      & 0.0     \\
NGC3838        &                           & 0.0      & 0.0      & 0.0      & 0.0      & 0.0      & 0.0      & 0.0      & 0.0      & 0.0     \\
NGC3941        & +wb                       & 0.0      & 0.0      & 0.0      & 0.0      & 0.0      & 0.0      & 1.0      & 0.0      & 0.0     \\
NGC3945        & +pb-pc                    & 0.0      & 0.0      & 0.0      & 0.0      & 0.0      & 0.0      & 2.0      & 0.0      & 2.0     \\
NGC3998        & +ah-h                     & 0.0      & 0.5      & 0.0      & 0.0      & 1.0      & 0.0      & 0.0      & 1.0      & 0.0     \\
NGC4026        & +wb                       & 0.0      & 0.0      & 0.0      & 0.0      & 0.0      & 0.0      & 1.0      & 0.0      & 0.0     \\
NGC4036        & +s+wl+wb                  & 2.0      & 0.0      & 0.0      & 0.0      & 0.0      & 1.0      & 1.0      & 0.0      & 0.0     \\
NGC4078        &                           & 0.0      & 0.0      & 0.0      & 0.0      & 0.0      & 0.0      & 0.0      & 0.0      & 0.0     \\
NGC4111        & +t-h                      & 0.0      & 0.0      & 2.0      & 0.0      & 0.0      & 0.0      & 0.0      & 1.0      & 0.0     \\
NGC4119        &                           & 0.0      & 0.0      & 0.0      & 0.0      & 0.0      & 0.0      & 0.0      & 0.0      & 0.0     \\
NGC4150        &                           & 0.0      & 0.0      & 0.0      & 0.0      & 0.0      & 0.5      & 0.0      & 0.0      & 0.0     \\
NGC4191        &                           & 0.0      & 0.0      & 0.0      & 0.0      & 0.0      & 0.0      & 0.0      & 0.0      & 0.0     \\
NGC4203        & +s?-h                     & 1.0      & 0.0      & 0.0      & 0.0      & 0.0      & 0.0      & 0.5      & 1.0      & 0.0     \\
NGC4215        & +pb                       & 0.0      & 0.0      & 0.0      & 0.0      & 0.0      & 0.0      & 2.0      & 0.0      & 0.0     \\
NGC4249        &                           & 0.0      & 0.0      & 0.0      & 0.0      & 0.0      & 0.0      & 0.0      & 0.0      & 0.0     \\
NGC4259        & +wb                       & 0.0      & 0.0      & 0.0      & 0.0      & 0.0      & 0.0      & 1.0      & 0.0      & 0.0     \\
NGC4261        & +r+ph-h                   & 0.0      & 2.0      & 0.0      & 0.0      & 2.0      & 0.0      & 0.0      & 1.0      & 0.0     \\
NGC4264        & +pb-h                     & 0.0      & 0.0      & 0.0      & 0.0      & 0.5      & 0.0      & 1.5      & 1.0      & 0.0     \\
NGC4268        & +pb                       & 0.0      & 0.0      & 0.0      & 0.0      & 0.5      & 0.0      & 2.0      & 0.0      & 0.0     \\
NGC4270        & +r+t+ph-h                 & 0.0      & 2.0      & 2.0      & 0.0      & 2.0      & 0.0      & 0.0      & 1.0      & 0.0     \\
NGC4278        & -h                        & 0.0      & 0.0      & 0.0      & 0.0      & 0.0      & 0.0      & 0.0      & 1.0      & 0.0     \\
NGC4281        & -h                        & 0.0      & 0.0      & 0.0      & 0.0      & 0.0      & 0.0      & 0.0      & 1.0      & 0.0     \\
NGC4283        & +r?+t?-h                  & 0.0      & 1.0      & 1.0      & 0.0      & 0.0      & 0.0      & 0.0      & 1.0      & 0.0     \\
NGC4382        & +s+r+ph-h-wc              & 2.0      & 2.0      & 0.0      & 0.0      & 2.0      & 0.0      & 0.0      & 1.0      & 1.0     \\
NGC4623        & +wb                       & 0.0      & 0.0      & 0.0      & 0.0      & 0.0      & 0.0      & 1.0      & 0.0      & 0.0     \\
NGC4624        & +ah+pb                    & 0.0      & 0.0      & 0.0      & 0.0      & 1.0      & 0.0      & 2.0      & 0.0      & 0.0     \\
NGC4636        & +ah                       & 0.0      & 0.0      & 0.0      & 0.0      & 1.0      & 0.0      & 0.0      & 0.0      & 0.0     \\
NGC4643        & +s?+t+ph+pb               & 1.0      & 0.0      & 1.5      & 0.0      & 2.0      & 0.0      & 2.0      & 0.0      & 0.0     \\
NGC4684        & +s+ph                     & 2.0      & 0.0      & 0.0      & 0.0      & 1.5      & 0.0      & 0.0      & 0.0      & 0.0     \\
NGC4690        & +r?+ph-h                  & 0.0      & 1.0      & 0.0      & 0.0      & 2.0      & 0.0      & 0.0      & 1.0      & 0.0     \\
NGC4697        & +ah                       & 0.0      & 0.0      & 0.0      & 0.0      & 1.0      & 0.0      & 0.0      & 0.0      & 0.0     \\
NGC4753        & +ph+pl                    & 0.0      & 0.0      & 0.0      & 0.0      & 2.0      & 2.0      & 0.0      & 0.0      & 0.0     \\
NGC5173        & +s?+r?+ah-h               & 1.0      & 1.0      & 0.0      & 0.0      & 1.0      & 0.0      & 0.0      & 1.0      & 0.0     \\
NGC5198        & +s+r                      & 2.0      & 2.0      & 0.0      & 0.0      & 0.0      & 0.0      & 0.0      & 0.0      & 0.0     \\
NGC5273        &                           & 0.0      & 0.5      & 0.0      & 0.0      & 0.0      & 0.0      & 0.0      & 0.0      & 0.0     \\
NGC5308        & +pb                       & 0.0      & 0.0      & 0.0      & 0.0      & 0.5      & 0.0      & 2.0      & 0.0      & 0.0     \\
NGC5322        & +ah                       & 0.0      & 0.0      & 0.0      & 0.0      & 1.0      & 0.0      & 0.0      & 0.0      & 0.0     \\
NGC5342        & +wb                       & 0.0      & 0.0      & 0.0      & 0.0      & 0.0      & 0.0      & 1.0      & 0.0      & 0.0     \\
NGC5353        & +t?+ph-h                  & 0.0      & 0.0      & 1.0      & 0.0      & 2.0      & 0.0      & 0.0      & 1.0      & 0.0     \\
NGC5355        & +s?+r?+t+ph-h             & 1.0      & 1.0      & 2.0      & 0.0      & 2.0      & 0.0      & 0.0      & 1.0      & 0.0     \\
NGC5358        & +pb-h                     & 0.0      & 0.0      & 0.0      & 0.0      & 0.0      & 0.0      & 1.5      & 1.0      & 0.0     \\
NGC5379        & +t+ah+wl+pb-h             & 0.0      & 0.0      & 2.0      & 0.0      & 1.0      & 1.0      & 2.0      & 1.0      & 0.0     \\
NGC5422        & +pl+wb                    & 0.0      & 0.0      & 0.0      & 0.0      & 0.0      & 2.0      & 1.0      & 0.0      & 0.0     \\
NGC5473        & +wb                       & 0.0      & 0.0      & 0.0      & 0.0      & 0.0      & 0.0      & 1.0      & 0.0      & 0.0     \\
NGC5481        & -h-wc                     & 0.0      & 0.0      & 0.0      & 0.0      & 0.0      & 0.0      & 0.0      & 1.0      & 1.0     \\
NGC5485        & +s+r?+ph+pl               & 2.0      & 1.0      & 0.0      & 0.0      & 2.0      & 2.0      & 0.0      & 0.0      & 0.0     \\
NGC5493        & +r+ph                     & 0.0      & 2.0      & 0.0      & 0.0      & 2.0      & 0.0      & 0.0      & 0.0      & 0.0     \\
NGC5500        &                           & 0.0      & 0.0      & 0.0      & 0.0      & 0.0      & 0.0      & 0.0      & 0.0      & 0.0     \\
NGC5507        & +r+t+ah+wb-h              & 0.0      & 2.0      & 2.0      & 0.0      & 1.0      & 0.0      & 1.0      & 1.0      & 0.0     \\
NGC5557        & +r+t+ph                   & 0.0      & 2.0      & 2.0      & 0.0      & 2.0      & 0.0      & 0.0      & 0.0      & 0.0     \\
NGC5574        & +t+ph+wb-h                & 0.0      & 0.0      & 2.0      & 0.0      & 2.0      & 0.0      & 1.0      & 1.0      & 0.0     \\
NGC5576        & +s+t?+ph-h                & 2.0      & 0.0      & 1.0      & 0.0      & 2.0      & 0.0      & 0.0      & 1.0      & 0.0     \\
NGC5582        & +d                        & 0.0      & 0.0      & 0.0      & 1.0      & 0.0      & 0.0      & 0.0      & 0.0      & 0.0     \\
NGC5611        & +ah+pb                    & 0.0      & 0.0      & 0.0      & 0.0      & 1.0      & 0.0      & 2.0      & 0.0      & 0.0     \\
NGC5631        & +s+r+ph+wl                & 2.0      & 2.0      & 0.0      & 0.0      & 2.0      & 1.0      & 0.0      & 0.0      & 0.0     \\
NGC5638        & +s+r-h                    & 2.0      & 2.0      & 0.0      & 0.0      & 0.5      & 0.0      & 0.0      & 1.0      & 0.0     \\
NGC5813        & +ah                       & 0.0      & 0.0      & 0.0      & 0.0      & 1.0      & 0.0      & 0.0      & 0.0      & 0.0     \\
NGC5831        &                           & 0.0      & 0.0      & 0.0      & 0.0      & 0.0      & 0.0      & 0.0      & 0.0      & 0.0     \\
NGC5838        & +pb                       & 0.0      & 0.0      & 0.0      & 0.0      & 0.0      & 0.0      & 2.0      & 0.0      & 0.0     \\
NGC5839        & +wb                       & 0.0      & 0.0      & 0.0      & 0.0      & 0.0      & 0.0      & 1.0      & 0.0      & 0.0     \\
NGC5845        & +ah-h                     & 0.0      & 0.0      & 0.0      & 0.0      & 1.0      & 0.0      & 0.0      & 1.0      & 0.0     \\
NGC5846        & -h                        & 0.0      & 0.0      & 0.0      & 0.0      & 0.5      & 0.0      & 0.0      & 1.0      & 0.0     \\
NGC5866        & +ph+pb                    & 0.0      & 0.0      & 0.0      & 0.0      & 2.0      & 0.0      & 2.0      & 0.0      & 0.0     \\
NGC6010        & +pb-pc                    & 0.0      & 0.0      & 0.0      & 0.0      & 0.0      & 0.0      & 2.0      & 0.0      & 2.0     \\
NGC6014        & +wl+wb                    & 0.0      & 0.0      & 0.0      & 0.0      & 0.0      & 1.0      & 1.0      & 0.0      & 0.5     \\
NGC6017        & +ah+wb                    & 0.0      & 0.0      & 0.0      & 0.0      & 1.0      & 0.0      & 1.0      & 0.0      & 0.0     \\
NGC6278        & +s?+wb-h                  & 1.0      & 0.0      & 0.0      & 0.0      & 0.0      & 0.0      & 1.0      & 1.0      & 0.0     \\
NGC6547        &                           & 0.0      & 0.0      & 0.0      & 0.0      & 0.0      & 0.0      & 0.0      & 0.0      & 0.0     \\
NGC6548        & +pb-pc                    & 0.0      & 0.0      & 0.0      & 0.0      & 0.0      & 0.0      & 2.0      & 0.0      & 2.0     \\
NGC6703        & -pc                       & 0.0      & 0.0      & 0.0      & 0.0      & 0.0      & 0.0      & 0.0      & 0.0      & 2.0     \\
NGC6798        & +pb-pc                    & 0.0      & 0.0      & 0.0      & 0.0      & 0.0      & 0.0      & 2.0      & 0.0      & 2.0     \\
NGC7280        & +s?+pb-h                  & 1.0      & 0.0      & 0.0      & 0.0      & 0.0      & 0.0      & 2.0      & 1.0      & 0.0     \\
NGC7332        & +ah+pb-h                  & 0.0      & 0.0      & 0.0      & 0.0      & 1.0      & 0.0      & 2.0      & 1.0      & 0.0     \\
NGC7454        & -pc                       & 0.0      & 0.0      & 0.0      & 0.0      & 0.0      & 0.0      & 0.0      & 0.0      & 2.0     \\
NGC7457        & -h-pc                     & 0.0      & 0.0      & 0.0      & 0.0      & 0.0      & 0.5      & 0.5      & 1.0      & 2.0     \\
NGC7465        & +t+d+ph+pl+pb-h-wc        & 0.0      & 0.0      & 2.0      & 1.0      & 2.0      & 2.0      & 2.0      & 1.0      & 1.0     \\
NGC7693        & +wb-h-pc                  & 0.0      & 0.0      & 0.0      & 0.0      & 0.5      & 0.0      & 1.0      & 1.0      & 2.0     \\
NGC7710        & +pl                       & 0.0      & 0.0      & 0.0      & 0.0      & 0.0      & 2.0      & 0.0      & 0.0      & 0.0     \\
PGC016060      & +d+pl-h                   & 0.0      & 0.0      & 0.0      & 1.0      & 0.0      & 2.0      & 0.0      & 1.0      & 0.0     \\
PGC028887      & +s?                       & 1.0      & 0.0      & 0.0      & 0.0      & 0.0      & 0.0      & 0.0      & 0.0      & 0.0     \\
PGC029321      &                           & 0.0      & 0.0      & 0.0      & 0.0      & 0.0      & 0.0      & 0.0      & 0.0      & 0.0     \\
PGC042549      & +pb                       & 0.0      & 0.0      & 0.0      & 0.0      & 0.0      & 0.5      & 2.0      & 0.0      & 0.0     \\
PGC050395      &                           & 0.0      & 0.0      & 0.0      & 0.0      & 0.0      & 0.0      & 0.0      & 0.0      & 0.0     \\
PGC056772      &                           & 0.0      & 0.0      & 0.0      & 0.0      & 0.0      & 0.0      & 0.0      & 0.0      & 0.0     \\
PGC058114      & +wl+wb                    & 0.0      & 0.0      & 0.0      & 0.0      & 0.0      & 1.0      & 1.0      & 0.0      & 0.0     \\
PGC061468      &                           & 0.0      & 0.0      & 0.0      & 0.0      & 0.0      & 0.0      & 0.0      & 0.0      & 0.0     \\
PGC071531      & -pc                       & 0.0      & 0.0      & 0.0      & 0.0      & 0.5      & 0.0      & 0.0      & 0.0      & 2.0     \\
UGC03960       &                           & 0.5      & 0.0      & 0.0      & 0.0      & 0.0      & 0.0      & 0.0      & 0.0      & 0.0     \\
UGC04551       &                           & 0.0      & 0.0      & 0.0      & 0.0      & 0.0      & 0.0      & 0.0      & 0.0      & 0.0     \\
UGC05408       & +s+r+ph                   & 2.0      & 2.0      & 0.0      & 0.0      & 2.0      & 0.0      & 0.0      & 0.0      & 0.0     \\
UGC06062       & +pb                       & 0.0      & 0.0      & 0.0      & 0.0      & 0.0      & 0.0      & 2.0      & 0.0      & 0.0     \\
UGC06176       & +pb                       & 0.0      & 0.0      & 0.0      & 0.0      & 0.0      & 0.0      & 2.0      & 0.0      & 0.0     \\
UGC08876       & +pb                       & 0.0      & 0.0      & 0.0      & 0.0      & 0.0      & 0.0      & 2.0      & 0.0      & 0.0     \\
UGC09519       & +d+wl                     & 0.0      & 0.0      & 0.0      & 1.0      & 0.0      & 1.0      & 0.0      & 0.0      & 0.0     \\

\end{longtable}

\bsp	
\label{lastpage}
\end{document}